\documentclass[11pt]{article} 
\newcommand{\version}{May, 2007}

\usepackage{amsthm,amsfonts,amsmath,amssymb}
\usepackage{graphicx}

\swapnumbers
\theoremstyle{plain}

\newtheorem{thm}{THEOREM}[section]
\newtheorem{lm}[thm]{LEMMA}

\theoremstyle{definition}

\theoremstyle{remark}
\newcommand{\upchi}{\raise1pt\hbox{$\chi$}}
\newcommand{\R}{{\mathord{\mathbb R}}}

\newcommand{\dd}{{\rm d}}

\renewcommand{\|}{{\Vert}}

\def\d{{\rm d}}

\numberwithin{equation}{section}  

\begin{document}

\markboth{\scriptsize{CGL \version}}{\scriptsize{CGL \version}}

\title{\bf{ Determination of the spectral gap in the Kac model for physical momentum and energy conserving collisions}}
\author{\vspace{5pt} Eric A. Carlen$^{1}$,  Jeffrey S. Geronimo$^{2}$, and Michael Loss$^{1}$ \\
\vspace{5pt}\small{ School of Mathematics, Georgia Institute of
Technology, Atlanta, GA, USA}
\\}
\date{\version}
\maketitle \footnotetext [1]{Work partially supported by U.S.
National Science Foundation
grant DMS 06-00037}
\maketitle \footnotetext [2]{Work partially supported by U.S.
National Science Foundation collaborative linkage grant.    \\
\copyright\, 2007 by the authors. This paper may be reproduced, in
its entirety, for non--commercial purposes.}

\begin{abstract}  The Kac model  
describes the local evolution of a gas of $N$ particles with {\it three} dimensional velocities by a random walk in which the steps correspond to binary collisions
that conserve momentum as well as energy. The state space of this walk is
a sphere of dimension $3N - 4$.  The Kac conjecture concerns the
spectral gap in the one step transition operator $Q$ for this walk. 
In this paper, we compute the exact spectral gap.

As in previous work by Carlen, Carvalho and Loss where a lower bound on the spectral gap was proved, we use a method that relates
the spectral properties of $Q$ to the spectral properties of a simpler operator $P$, which is simply an average of certain non--commuting projections.   The new feature is that we show how to
use a knowledge of certain eigenfunctions and eigenvalues of $P$ to determine spectral properties of $Q$, instead of simply using the spectral gap for $P$ to bound the spectral gap for $Q$, inductively in $N$, as in previous work.  The methods developed here can be applied to many other high--dimensional stochastic process, as we shall explain.

We also use some deep results on Jacobi polynomials to obtain the required spectral information on $P$,  and we show how the identity through which Jacobi polynomials enter our problem may be used to obtain new bounds on Jacobi polynomials.

\end{abstract}


\newpage

\section{The Markov transition operator  $Q$ for the Kac walk} \label{Q}
\medskip

Let $X_N$ be the $N$ particle state space
consisting of $N$--tuples  $\vec v = (v_1,\dots , v_N)$ of vectors $v_j$ in $\R^3$ with
$$\sum_{j=1}^N|v_j|^2 =1\qquad{\rm and}\qquad   \sum_{j=1}^N v_j = 0\ .$$
We think of a point $\vec v$ as specifying the velocities of $N$ particles, and shall consider a random walk on $X_N$ that was  introduced by Mark Kac \cite{K}.  At each step of this random walk, $\vec v$ is updated  due to the effect of a binary collision that conserves energy and momentum  --- hence the constraints defining $X_N$. 

To specify this walk in more detail, we consider a collision in which  particles
$i$ and $j$ collide. Suppose that   $v_i^*$ and $v_j^*$ are the post--collisional velocities, while $v_i$ and $v_j$ are the pre--collisional velocities. Then by momentum conservation,
the center of mass velocity is conserved; i.e.,
$$v_i^*+v_j^*  = v_i + v_j\ .$$
Furthermore,  by energy conservation, i.e., $|v_i^*|^2+|v_j^*|^2 = |v_i|^2+|v_j|^2$, and the parallelogram law,
it follows that
$$|v_i^*-v_j^*|  = |v_i - v_j|\ .$$
This leads to a natural parameterization of all the possible binary collision outcomes that conserve energy and momentum: The parameter $\sigma$ is a unit vector in $S^2$, and when particles
$i$ and $j$ collide, one updates
$\vec v \to \vec v^* = R_{i,j,\sigma}(\vec v)$ where
\begin{eqnarray}\label{crules}
v_i^* &=&  \frac{v_i+v_j}{2} +   \frac{|v_i -v_j|}{2}\sigma \nonumber\\
v_j^* &=&  \frac{v_i+v_j}{2} -   \frac{|v_i -v_j|}{2} \sigma  \nonumber\\
v_k^* & =&v_k\quad{\rm for}\ k \ne i,j\ . \nonumber\\
\end{eqnarray}

The {\it Kac walk} on $X_N$ is a random walk in which the steps are such binary collisions 
between pairs of particles.  At each step, one picks a pair $(i,j)$, $i<j$ uniformly at random, and also  a unit vector $\sigma$ in $S^2$. One then makes the update described in (\ref{crules}).  Of course it remains to specify the probabilistic rule according to which $\sigma$ should be selected. 
In the physics being modeled here, the likelihood of selecting a particular $\sigma$ will depend only on the resulting {\em scattering angle} $\theta$ , which is the angle between $v_i^*-v_j^*$ and $v_i-v_j$. 
In the parameterization above, this is the angle between $\sigma$ and  $v_i-v_j$. That is,
$$\cos(\theta)=  \sigma\cdot \frac{v_i-v_j}{|v_i-v_j|}\ .$$
The {\em scattering rate function} $b$ is a  non negative integrable  function on $[-1,1]$ with
$$\frac{1}{2}\int_{-1}^1 b(u)\d u =1\ .$$ Then for any $v_i \ne v_j$, and with $\d \sigma$ being the uniform probability measure on $S^2$,
\begin{equation}\label{den}
\int_{S^2}b\left(\sigma\cdot \frac{v_i-v_j}{|v_i-v_j|}\right)\dd \sigma =1\ .
\end{equation}
(If $v_i =v_j$, the collision has no effect, and can be ignored.) 
One selects $\sigma$  according to the probability density that is integrated in (\ref{den}). 

There are several choices of $b$ of particular interest. One is the {\it uniform redirection model}, given by $b(x) =1$ for all $-1\le x\le 1$. In this case, the new direction of the relative velocity, $\sigma$ is chosen uniformly from $S^2$. 

Another is the {\it Morgenstern model} \cite{Mo54},\cite{Mo55}, or the {\it uniform reflection model}:  For any unit vector $\omega\in S^2$, let
$H_\omega$ be the reflection given by
$$H_\omega(v) = v - 2(v\cdot\omega)\omega\ .$$
In the uniform reflection model, one updates the relative velocity according to
$$v_i-v_j  \to H_\omega(v_i-v_j) = v_i^* - v_j^*$$
with $\omega$ chosen uniformly. The relation between $\omega$ and $\sigma$ is given by
$\sigma = H_\omega((v_i-v_j)/|v_i-v_j|)$, and computing the Jacobian of the map $\omega \to \sigma$, one finds
$$b(x) = \frac{1}{\sqrt{2}\sqrt{1- x}}\ .$$

Both of these belong to the one parameter family
\begin{equation}\label{balpha}
b_\alpha(x) = (1-\alpha)2^{\alpha}(1-x)^{-\alpha}\ .
\end{equation}
Leaving the particular choice of $b$ open, this completes the specification of the steps in the Kac walk. For more detail and background, see \cite{K} and  \cite{CCL2}.

The main object of study here is the spectrum of the one step  transition operator  $Q$ for this random walk, and the manner in which this spectrum depends on $N$.   $Q$ is defined as follows:
Let $\vec v_n $ be state of the process after the $n$th step. 
The one step Markov transition operator  $Q$ is given by taking the conditional expectation
$$Q\phi(\vec v) = {\rm E}(\phi(\vec v_{n+1})\ |\ \vec v_n = \vec v)\ ,$$
for any continuous function $\phi$ on $X_N$. 

From the above description, one deduces the formula
\begin{equation}\label{qopdef}
Q\phi(\vec v) =\left(
\begin{array}{c}
N\\ 2\end{array} \right)^{-1}\sum_{i<j}\int_{S^2}\phi(R_{i,j,\sigma}(\vec v))
b\left(\sigma\cdot \frac{v_i-v_j}{|v_i-v_j|}\right){\rm d}\sigma\ .
\end{equation}

Let $\sigma_N$ denote the uniform probability measure on $X_N$, which is the 
normalized measure
induced on $X_N$ as a manifold embedded in $\R^{3N}$.  

For any two unit vectors $\sigma$ and $\omega$, one sees from (\ref{crules}) that 
$$R_{i,j,\sigma}(R_{i,j,\omega}\vec v) =  R_{i,j,\sigma}\vec v\ .$$
From this and the fact that the measure $\dd \sigma_N$ is invariant under $\vec v \mapsto R_{i,j,\sigma}\vec v$, it follows that for any continuous functions $\phi$ and $\psi$ on $X_N$,
$$\int_{X_N}\psi (\vec v)Q\phi(\vec v)\dd \sigma_N  =
\int_{X_N}\int_{S^2}\int_{S^2}\psi(R_{i,j,\omega}\vec v)\phi(R_{i,j,\sigma}\vec v)b(\omega\cdot\sigma)
\dd \omega \dd \sigma \dd \sigma_N\ .$$
It follows that  $Q$ is a self adjoint Markov operator on $L^2(X_N,\sigma_N)$. Moreover, it is clearly a Markov operator; that is,
 in addition to being self adjoint, $Q$ is positivity preserving and $Q1 = 1$.

The motivation for considering the spectral properties of $Q$ stems from a theorem of Kac \cite{K}
that relates the continuous time version of the Kac walk to the nonlinear Boltzmann equation.
For the details, see \cite{K} or \cite{CCL2}. Let $\vec v(t)$ denote the random variable giving the state
of the system at time $t$ for the process run in continuous time with the jumps taking place in a Poisson stream with the mean time between jumps being $1/N$.
Then the equation describing the evolution 
of the probability law of  $\vec v(t)$, 
is called the {\it Kac Master Equation}:    If the initial law on $X_N$ has a density $F_0$, then
the law at time $t$ has a density $F(\vec v,t)$  satisfying
$${\partial \over \partial t}F(\vec v,t) = N(I -Q)F(\vec v,t)\qquad {\rm with}\qquad
F(\vec v,0) = F_0(\vec v)\ .$$
The solution $F(\vec v,t)$ is of course given by
$$F(\vec v,t) = e^{t{\cal L}}F_0(\vec v)\ ,$$
where
$${\cal L} = N(Q-I)\ .$$

Since $Q$ is a self adjoint Markov operator, its spectrum lies in the interval $[-1,1]$, and since $Q1= 1$,  the constant function is an eigenfunction with eigenvalue $1$. It is easily seen that as long as $b(x)$ is strictly positive on a neighborhood of
$x=1$, the   eigenvalue $1$ of $Q$ has multiplicity one. It then  follows that the  spectrum of ${\cal L}$
lies in $[-2N,0]$, and that $0$ is an eigenvalue of multiplicity one.  We impose this assumption on $b$ throughout what follows.

The {Kac conjecture} for this stochastic process pertains to the spectral gap
$$\Delta_N = \inf\left\{-\int_{X_N}\phi(\vec v) {\cal L}\phi(\vec v){\rm d}\sigma_N \ \bigg|\  
\int_{X_N}\phi^2(\vec v){\rm d}\sigma_N = 1\ ,\ \int_{X_N}\phi(\vec v){\rm d}\sigma_N = 0\ \right\}\ ,$$
and states that
$$\liminf_{N\to\infty}\Delta_N > 0\ .$$

This was proved by Carlen, Carvalho and Loss \cite{CCL2}, but without an explicit lower bound.    Kac also made a similar conjecture for a simplified model with one dimensional velocities and no conservation of momentum.  For this model,
the conjecture was first proved by Janvresse \cite{J}, though her approach provided no explicit lower bound. The sharp bound
for the simplified model was first established in  \cite{CCL1}. See  Maslen \cite{Mas} for a representation theoretic approach.

The main goal of the present paper is to compute {\it exactly}  $\liminf_{N\to\infty}\Delta_N$.  We shall be able to do this under an easily checked condition
relating $\Delta_2$ and the quantities
\begin{equation}\label{b2def}
 B_1 = \frac{1}{2}\int_{-1}^1 x b(x)\d x  \qquad{\rm and}\qquad  B_2 = \frac{1}{2}\int_{-1}^1 x^2 b(x)\d x\ .
\end{equation}
 The condition, given in (\ref{d2cond}) below,  will turn out to be satisfied  when $b$ is given by $b_\alpha$, as in (\ref{balpha}),
for all $0\le \alpha \le 7/9$.

\medskip

\begin{thm}\label{exact} Suppose  that  $B_2> B_1$ and that
\begin{equation}\label{d2cond}
\Delta_2 \ge  \frac{20}{9}(1-B_2)\ .
\end{equation}
Then for all $N\ge 3$, 
\begin{equation}\label{gap1}
\Delta_N = (1-B_2)\frac{N}{(N-1)}\ .\end{equation}

Moreover,
the eigenspace is three dimensional, and is spanned by the functions
\begin{equation}\label{eig1}
\phi(\vec v) = \sum_{j=1}^N |v_j|^2v_j^\alpha\qquad \alpha =1,2,3\ ,
\end{equation}
where $v_j^\alpha$ denotes the $\alpha$th component of $v_j$. 
\end{thm}

\medskip

As we shall see in the next section, for many choices of $b$, including the $b_\alpha$, there is a simple monotonicity of the eigenvalues of $Q$ for $N=2$ which ensures that the eigenfunction providing the gap comes from a first degree polynomial, and thus that
\begin{equation}\label{d2form}
\Delta_2 = 2(1- B_1)\ .
\end{equation}
When (\ref{d2form}) is satisfied,
 the condition (\ref{d2cond}) reduces to $(1-B_1)/(1-B_2) > 20/9$. 

Next, notice that the eigenfunctions listed in (\ref{eig1}) are symmetric under permutation of the particle indices. Indeed, the operator $Q$ commutes with such permutations, so that the subspace of functions with this symmetry is invariant. As explained in \cite{K} and \cite{CCL2}, it is the spectrum of $Q$ on this subspace that is relevant for the study of the Boltzmann equation. 

Moreover, notice that  in the collision rules (\ref{crules}),  exchanging $v_i^*$ and $v_j^*$
has the same effect as changing $\sigma$ to $-\sigma$. For this reason, if one's primary object of interest is the Boltzmann equation, one may freely assume that $b$ is a symmetric function on $[-1,1]$,
since then replacing $b(x)$ by $(b(x) + b(-x))/2$ will have no effect on the spectrum of $Q$ on the invariant subspace of symmetric functions, or on the corresponding Boltzmann equation. (See the introduction
 of \cite{CL} for more discussion of this point in the context of the Boltzmann equation.)  If $B$ is symmetric, then $B_1 = 0$, and we do have $B_1 >  B_2$. 
 
However, it is interesting that the Kac conjecture holds without restriction to the symmetric subspace,
and the that methods developed here can be used to determine the spectral gap even when $b$
is not symmetric, and the eigenfunctions corresponding to the gap eigenvalue are not symmetric.

When $b$ is not symmetric, it may happen that $B_1 \le B_2$.  We shall give examples of
this below. The next theorem gives the spectral gap and the eigenfunctions whenever
$\Delta_2 = 2(1-B_1)$, regardless of whether $B_1<B_2$ or $B_2<B_1$. However, it gives the exact value of $\Delta_N$ only for $N\ge 7$. Since we are interested in large values of $N$, this is fully satisfactory. Indeed, it is remarkable that the two theorems show that already at relatively small values of $N$, the behavior of the system is very close, qualitatively and quantitatively to the behavior in the large $N$ limit.

\vfill\break

\begin{thm}\label{exact2} Suppose  that  
$\Delta_2 = 2(1-B_1)$.
Then for all $N\ge 7$, 
\begin{equation}\label{gap1b}
\Delta_N = \min\{\ (1-B_1)\ ,\ (1-B_2)\ \}\frac{N}{(N-1)}\ .
\end{equation}

Moreover, if $B_2> B_1$,
the eigenspace is three dimensional, and is spanned by the functions
\begin{equation}\label{eig1b}
\phi(\vec v) = \sum_{j=1}^N |v_j|^2v_j^\alpha\qquad \alpha =1,2,3\ ,
\end{equation}
where $v_j^\alpha$ denotes the $\alpha$th component of $v_j$. 

On the other hand, if  $B_2<B_1$, the eigenspace is spanned by the functions of the form
\begin{equation}\label{eig2}
|v_i|^2 - |v_j|^2 \qquad{\rm and}\qquad v^\alpha_i - v^\alpha_j\ , \quad  \alpha =1,2,3\ ,
\end{equation}
for all $i<j$. 

Finally, if $B_1 = B_2$, the eigenspace is spanned by both of the sets of functions listed in (\ref{eig1}) and (\ref{eig2})
together. 
\end{thm}

\medskip 
For the family of collision rates introduced so far, the $b_\alpha$, one may apply Theorem \ref{exact}, as we have indicated, but only for
$\alpha \le 7/9$.   As we shall see in Section 2,   Theorem \ref{exact2} applies for all  $0\le  \alpha < 1$, and in this case gives
$\Delta_N = (N/N-1)(1-B_2)$ for $N\ge 7$.  In order to illustrate the case in which Theorem \ref{exact2} yields the gap  $\Delta_N = (N/N-1)(1-B_1)$,
we introduce
\begin{equation}\label{btilde}
\widetilde b_\alpha(x) = 2(\alpha +1)1_{[0,1]}(x) x^\alpha\qquad \alpha \ge 0\ .
\end{equation}
Since $x^2 < x$ on $(0,1)$, it is clear that $B_2 < B_1$ for all $\alpha$ in this case. We show  at the end of Section 2 that at least for 
$0 \le \alpha \le 1$, $\Delta_2 = 2(1-B_1)$, so that Theorem \ref{exact2} applies in in these cases.

\medskip

The method of proof is quite robust, and 
in Section 10 we shall describe how it may be extended to determine the spectral
gap of $Q$ for still other choices of $b$ that are not covered by the Theroems \ref{exact} and \ref{exact2}.

The method of proof of these theorems  relies on a basic strategy introduced in \cite{CCL2}, 
but which is extended significantly here. The strategy consists of exploiting an inductive link between the spectral gap of $Q$ and the one of an operator $P$, an average over
projections introduced in Section 3. In fact, 
\begin{equation}\label{indl}
\Delta_N \ge { N \over N-1} (1-\mu_N) \Delta_{N-1}  
\end{equation}
where $1-\mu_N$ is the gap of $P$.
The eigenvalues of $P$ are much easier to compute than the ones of $Q$ since the range of $P$ consists of sums of functions of single variables $v_j$.

In the case of the original model treated by Kac, one is in the happy circumstance that
$Q$ has a single gap eigenfunction $\phi$ which is also the gap eigenfunction of $P$
for all $N$, {\it and} when this is used as a trial function in the derivation of (\ref{indl}), one
sees that (\ref{indl}) actually holds with equality, giving an identity relating $\Delta_N$ and $\Delta_{N-1}$. Thus, starting at $N=2$, where the gap can be easily
calculated, the above formula yields a lower bound on $\Delta_N$ that
turns out to be exact.
The model treated in this paper does not have this simplifying feature, even when the gap eigenfunctions of $Q$ are also the gap eigenfunctions of $P$. Nevertheless, the  ideas that lead to (\ref{indl})
can be used in such a way that we can still calculate the gap of $Q$ exactly. Very briefly, here is how:

Let $\mu_N^*<\mu_N$
be any number and  assume that
there are finitely many eigenvalues $\mu_N^* \le \mu_N^{(m)}\le
\cdots \le \mu_N^{(1)} \le \mu_N$ of $P$. Denote the corresponding eigenspaces
by $E_j$. Let $V_j$ be the smallest invariant subspace of $Q$
that contains $E_j$.  Lemma 4.1 in Section 4 provides us with the following dichotomy:  
{\it Either}
\begin{equation}\label{indl2}
\Delta_N \ge { N \over N-1} (1-\mu ^*_N) \Delta_{N-1} 
\end{equation}
{\it or else:}  
\begin{equation}\label{indl5}
{\rm The\ gap\ of}\ Q\ {\rm is\ the\ same\ as\ the\ gap\ of }\ Q\  {\rm restricted\
to}\ \oplus_{j=1}^m V_j.  
\end{equation}
If the threshold has been chosen so that the lower bound on $\Delta_N$
provided by  (\ref{indl2}) is at least as large as the upper bound on $\Delta_N$ provided by some trial function in 
$\oplus_{j=1}^m V_j$, then $\Delta_N$ is 
 the gap of $Q$ restricted to $\oplus_{j=1}^m V_j$. As we shall see, the $V_j$ are  finite dimensional, so determining the gap of $Q$ on $\oplus_{j=1}^m V_j$ is a tractable problem. In this case we have determined the exact value of $\Delta_N$.
 
To proceed to the determination of $\Delta_N$ for all large $N$, one needs a strategy for choosing the threshold $\mu_N^\star$. The lower the value of $\mu_N^\star$ that is chosen, the stronger the bound
(\ref{indl2}) will be, but also the higher the value of $m$ will be.  The basis for the choice of $\mu_N^\star$ is a trial function calculation, providing a guess $\widetilde\Delta_N$ for the value of $\Delta_N$.
Indeed, natural trial functions can often be chosen on the basis of physical considerations. (The spectrum of the linearized Boltzmann equation is the source in the case at hand.) To show that the guess is correct, so that $\widetilde\Delta_N=\Delta_N$, we are led to choose $\mu_N^\star$ so that
\begin{equation}\label{indl3}
\widetilde\Delta_N \le { N \over N-1} (1-\mu ^*_N) \widetilde\Delta_{N-1} 
\end{equation}
Since $\widetilde\Delta_{N-1} \ge \Delta_{N-1}$, this forces us into the second alternative in the dichotomy discussed above, so that the gap eigenfunction for $N$ particles lies in 
$\oplus_{j=1}^m V_j$.  Indeed, if the physical intuition behind the guess was correct, the trial function leading to $\widetilde \Delta_N$ will lie in $\oplus_{j=1}^m V_j$, and yield the gap.

Choosing $\mu_N^\star$ small enough that (\ref{indl3}) is satisfied might in principle lead to a value of $m$ that depends on $N$. However, in the case at hand, we are fortunate, and can work with a choice of $\mu_N^\star$ that leads to a fixed and small value of $m$, but for which 
(\ref{indl3}) is satisfied for all sufficiently large values of $N$ -- hence the restriction to $N\ge 7$ in Theorem \ref{exact2}.

As will be clear from this summary of the strategy, the determination of the spectrum of $P$  is
the main technical step that must be accomplished. As we mentioned before, this is relatively simple, compared to the determination of the spectrum of $Q$, since the range of $P$ consists of functions that are a sum of functions of a single variable. 

For this reason, we can reduce the study of the spectrum of $P$ to that of 
 a much simpler
Markov operator $K$ acting on functions on the unit ball $B$ in $\R^3$.
In the analysis of 
$K$,
we shall  draw on some deep results on Jacobi polynomials \cite{Kor},\cite{NEM}. In fact,   it turns out that the  connection between our eigenvalue  problems and pointwise bounds on  Jacobi
polynomials is through a simple identity, and applications of this identity can be made in both directions: We not only use bounds on Jacobi polynomials to bound eigenvalues, we
shall use simple eigenvalue estimates to sharpen certain  best  known bounds
on Jacobi polynomials, as we briefly discuss in Section 11.

First however, we deal with a simpler technical problem, the 
computation of the spectral gap of $Q$ for $N=2$.  

\medskip
\section{The spectral gap for $N=2$} \label{N2}
\medskip

For $N=2$, the state space $X_2$ consists of pairs $(v,-v)$ with $v\in \R^3$ satisfying $|v|^2 = 1/2$. 
Note that for $N=2$  the collision rules (\ref{crules}) reduce to
$$v_1^* = \sigma/\sqrt{2}  \qquad{\rm and}\qquad   v_2^* =-\sigma/\sqrt{2}\ ,$$
since $v_1+v_2 =0$. 

The map
$(v,-v)\mapsto \sqrt{2}v$ identifies $X_2$ with the unit sphere $S^2$, and the measure 
$\d \sigma_2$ on $X_2$ with $\d \sigma$ on $S^2$. Thus, we may think of $Q$ as operating on functions on $S^2$.
In this representation, we have the formula
$$Q\phi(u) = \int_{S^2}\phi(\sigma)b(u\cdot\sigma)\d \sigma \ .$$

Notice that if $R$ is any  rotation of $\R^3$ 
$$(Q\phi)(Ru) =  \int_{S^2}\phi(\sigma)b(Ru\cdot\sigma)\d \sigma =  \int_{S^2}\phi(R\sigma)b(Ru\cdot R\sigma)\d \sigma
=  \int_{S^2}\phi(R\sigma)b(u\cdot \sigma)\d \sigma = Q(\phi\circ R)(u)\ . $$
 That is, $(Q\phi)\circ R = Q(\phi\circ R)$, and this means that for each $n$, the space of spherical harmonics of degree $n$
 is an invariant subspace of $Q$, contained in an eigenspace of $Q$. In turn, this means that we can determine the spectrum of $Q$ by computing its action on the zonal spherical harmonics, i.e., those of the form $P_n(e\cdot u)$ where $e$ is any fixed
 unit vector in $\R^3$, and $P_n$ is the $n$th degree Legendre polynomial.  Now, for any function $\phi(u)$ of the form
 $\phi(u) = f(e\cdot u)$, 
 $$Q\phi(u) = \int_{S^2} \phi(\sigma\cdot e)b(\sigma\cdot u)\d \sigma\ .$$ We choose coordinates in which
 $u$ and $e$ span the $x,z$ plane with $u = \left[
 \begin{array}{c}
0\\ 0\\1\end{array}\right]$ and $e = \left[ \begin{array}{c} \sqrt{1-t^2}\\ 0\\ t \end{array}\right]$,
 so that $t= u\cdot e$. 
 Then with $\sigma = \left[\begin{array}{c} 
 \sin\theta\sin\varphi\\ 
 \cos\theta\sin\varphi\\ 
 \cos\theta\end{array}
 \right]$,
 $Q\phi(u) = {\cal Q}f(e\cdot u)$ where
\begin{eqnarray}
{\cal Q}f(t) &=& \frac{1}{4\pi}\int_0^\pi\int_0^{2\pi} f(t\cos\theta + \sqrt{1-t^2}\sin\theta\cos\varphi)b(\cos\theta)\sin\theta\d \theta \d \varphi\nonumber\\
&=& \frac{1}{4\pi}\int_0^\pi\int_{-1}^{1} f(ts + \sqrt{1-t^2}\sqrt{1-s^2}\cos\varphi)b(s)\d s \d \varphi\ .\nonumber\\
\end{eqnarray}
Now, if $f$ is any eigenfunction of ${\cal Q}$ with ${\cal Q}f = \lambda f$, then evaluating both sides at $t=1$, we have
${\displaystyle \lambda f(1) = \frac{1}{2}\int_0^\pi\int_{-1}^{1} f(s)b(s)\d s}$.
Thus, the eigenvalue is given by
$$\lambda = \frac{1}{2}\int_{-1}^{1} \frac{f(s)}{f(1)}b(s)\d s\ .$$
As we have observed above, the eigenfunctions of ${\cal Q}$ are the Legendre polynomials. Thus, if $P_n$
is the  Legendre polynomial of $n$th degree with the standard normalization $P_n(1) =1$, and $\lambda_n$ is the corresponding eigenvalue,
\begin{equation}\label{eigform}
\lambda_n = \frac{1}{2}\int_{-1}^{1} P_n(s)b(s)\d s\ .
\end{equation}

This explicit formula enables one to easily compute $\Delta_2$.  For example, we can now easily prove the following:

\begin{lm}\label{gapba}
When $b(x) =  b_\alpha (x)$, as in (\ref{balpha}), then $1-B_2 < 1-B_1$ for all $\alpha<1$, and moreover, 
 \begin{equation}\label{del2}
 \Delta_2 = 2(1- \lambda_1)  = \frac{4(1-\alpha)}{{2-\alpha}}  = (1- B_2)(3-\alpha)\ ,
 \end{equation}
 so that (\ref{d2cond}) is satisfied for all $\alpha$ with $0 \le \alpha \le 7/9$.
\end{lm}

\noindent{\bf Proof:} 
Using Rodrigues' formula
 $$P_n(x) = \frac{1}{2^n n!} \frac{{\rm d}^n}{{\rm d}x^n}(x^2-1)$$
 and integration by parts, one computes
 $$\lambda_n =   (1-\alpha)\frac{(\alpha)_n}{(1-\alpha)_{n+1}} = \frac{(\alpha)_n}{(2-\alpha)_n}\ ,$$
 where $(\alpha)_n = \alpha(\alpha+1)(\alpha+2) \cdots (\alpha+n-1)$.  
 Notice that for all $0 \le \alpha < 1$,  $\lambda_n$ decreases as $n$ increases, so with the collision rate given by $b_\alpha$,
 \begin{equation}\label{del2b}
 \Delta_2 = 2(1- \lambda_1)  = \frac{4(1-\alpha)}{{2-\alpha}}\ .
 \end{equation}
 
 Next, one computes
 $$1-B_1 =   \frac{2(1-\alpha)}{(2-\alpha)}  \qquad{\rm and}\qquad  1 - B_2 =  \frac{4(1-\alpha)}{(2-\alpha)(3-\alpha)}\ .$$
 Since $2> 4/(3-\alpha)$ for $\alpha<1$, $1-B_2 < 1-B_1$ for all $\alpha < 1$.  Moreover, from this computation,
 one readily obtains (\ref{del2}) and the statement concerning (\ref{d2cond}). \qed
 \medskip
 
 In particular, the condition   (\ref{d2cond}) is satisfied in both the uniform redirection model ($\alpha =0$) and
 the Morgenstern model ($\alpha =1/2$). Thus in these cases we have the exact spectral gaps
 \begin{eqnarray}
 \Delta_N &=& \frac{2}{3}\frac{N}{N-1}\qquad {\rm for\ the\   uniform\ redirection\ model}\nonumber\\
 \Delta_N &=& \frac{8}{15}\frac{N}{N-1}\qquad {\rm for\ the\   Morgenstern\ model}\nonumber\\
 \end{eqnarray}
 
We close this section with a remark that may provide a useful perspective on what follows. In determining the spectral
gap of $Q$ for $N=2$, general symmetry conditions told us right away what all of the eigenfunctions were. A less
obvious, though still simple, argument then provided us with the explicit formula (\ref{eigform}) for all of the eigenvalues.
There is one last hurdle to cross: There are infinitely many eigenvalues given by (\ref{eigform}), and for a general $b$,
we cannot determine which is the second largest by computing them all explicitly. What was particularly nice about $b_\alpha$
is that in this case, the eigenvalues of $Q$ were monotone decreasing:
$$\lambda_{n+1} \le \lambda_n\ .$$
For other choices of $b$, this need not be the case. However, there are ways to use {pointwise bounds} on Legendre polynomials to reduce the problem of determining $\Delta_2$
 to the computation of a {\it finite} number of eigenvalues using  (\ref{eigform}).  For example, one has
 the classical bound (see Theorem 7.3.3 in \cite{Szego}):
 \begin{equation}\label{polya}
 |P_n(x)|^2 < \frac{2}{n\pi}\frac{1}{\sqrt{1-x^2}}\ .
 \end{equation}
 As long as $b(x)(1-x^2)^{-1/4}$ is integrable, this gives an upper bound on $\lambda_n$ that is proportional to $n^{-1/2}$:
 Define 
 $$\tilde \lambda_n= \left(\frac{1}{8\pi n}\right)^{1/2}\int_{-1}^1 b(x) (1-x^2)^{-1/4}\d x \ .$$
 Then, let $n_0$ be the least value of $n$ such that $\tilde \lambda_n \le \lambda_1$.  Then the second largest eigenvalue of 
 $Q$ is
 $$\max_{1 \le n \le n_0} \lambda_n\ .$$

 We illustrate this by showing that for the rate function  $\widetilde b_\alpha$ introduced in (\ref{btild}), $\Delta_2 =2(1- B_1)$
 at least for $0 \le \alpha \le 1$.  (Of course, the integrals in (\ref{eigform}) can be computed exactly in this case; see $7.231$, page 822 in \cite{GR}. however, we prefer to illustrate the use of (\ref{polya})).

By (\ref{polya}) and (\ref{eigform}),
\begin{eqnarray}
|\lambda_n| &\le& (\alpha+1)\left(\int_0^1 x^{2\alpha}\d x\right)^{1/2}\left(\int_0^1P_n(x)^2\d x \right)^{1/2}\nonumber\\
&<& \frac{\alpha+1}{\sqrt{2\alpha+1}}\frac{1}{\sqrt{n}}\ .\nonumber\\
\end{eqnarray}

Also, by (\ref{eigform}), $\lambda_1 = B_1 = (\alpha+1)/(\alpha +2)$. Comparison  of the formulas shows that for  $0 \le \alpha \le 1$,
$\lambda_n < \lambda_1$ for all $n > 4$.  Thus it suffices to check that $\lambda_j < \lambda_1$ for $j=2,3$ and $4$ by direct computation with (\ref{eigform}). Doing so, one finds that this is the case.  Hence, Theorem \ref{exact2} applies, and yields 
$\Delta_N = (N/N-1)(1-B_1)$ for $N \ge 7$.

Further calculation would extend this result to higher values of $\alpha$. Notice that as $\alpha$ tends to infinity, 
$\widetilde b_\alpha(x)$ is more and more concentrated at $x=1$, which corresponds to $\theta = 0$. Thus, for large values of
$\alpha$, $\widetilde b_\alpha$ represents a ``grazing collision model''.

For $N>2$, the operator $Q$ is much more complicated, and direct
determination of the spectrum is not feasible. Instead, we use an inductive procedure involving a auxiliary operator that we now introduce.

\medskip
\section{The average of projections operator $P$, and its relation to $Q$} \label{P}
\medskip

A simple convexity argument shows that for each $j$,
$$\sup\{|v_j|^2 \ :\ \vec v \in X_N\} = {N-1\over N}\ .$$
For each $j$, define $\pi_j(\vec v)$ by
$$\pi_j(\vec v) = \sqrt{{N\over N-1}}v_j\ ,$$
so that $\pi_j$ maps $X_N$ onto the unit ball $B$ in $\R^3$. 

For any function $\phi$ in $L^2(X_N,{\rm d}\sigma_N)$, and any $j$ with $1\le j \le N$, define
$P_j(\phi)$ to be the orthogonal projection of $\phi$ onto the subspace of 
 $L^2(X_N,{\rm d}\sigma_N)$ consisting of square integrable functions that depend on $\vec v$ through $v_j$ alone.   That is,  $P_j(\phi)$ is the unique element of  $L^2(X_N,{\rm d}\sigma_N)$
of the form $f(\pi_j(\vec v))$ such that
$$\int_{X_N}\phi(\vec v)g(\pi_j(\vec v)){\rm d}\sigma_N = 
\int_{X_N}f(\pi_j(\vec v))g(\pi_j(\vec v)){\rm d}\sigma_N $$
for all continuous functions $g$ on $B$. 

The {\it average of projections operator} $P$ is then defined through
$$P = {1\over N}\sum_{j=1}^N P_j\ .$$
 
 If the individual projections $P_j$ all commuted with one another, then the spectrum
 of $P$ would be very simple: The eigenvalues of each $P_j$ are $0$ and $1$.
 Moreover, $P_j\phi = \phi$ if and only if $\phi$ depends only on $v_j$ so that it cannot
 then also satisfy $P_k\phi = \phi$ for $k\ne j$, unless $\phi$ is constant. It would then follow that the eigenvalues of $P$
would be  $0$, $1/N$  and $1$, with the last having multiplicity one. 

However, the individual projections $P_j$ do not  commute with one another, due to the nature of the constraints defining $X_N$.  

We now define
\begin{equation}\label{pgap}
\mu_N = \sup\left\{\int_{X_N}\phi(\vec v) P \phi(\vec v){\rm d}\sigma_N \ \bigg|\  
\int_{X_N}\phi^2(\vec v){\rm d}\sigma_N = 1\ ,\ \int_{X_N}\phi(\vec v){\rm d}\sigma_N = 0\ \right\}\ .
\end{equation}
The $P$ operator is simpler than the $Q$ operator in that if $\phi$ is any eigenfunction of $P$ with non--zero eigenvalue,
then clearly $\phi$ has the form
$$\phi = \sum_{j=1}^Nf_j\circ \pi_j\ $$
for some functions $f_1,\dots,f_N$ on $B$.
For $N\ge 4$, most of the eigenfunctions of $Q$ have a more complicated structure. 
Nonetheless, there is a close relation between the spectra of $Q$ and $P$, as we now explain.

To do this, we need a more explicit formula for $P$, such as the formula (\ref{qopdef}) that we have for $Q$.
The key to computing $P_j\phi$ is a factorization formula \cite{CCL2} for the measure 
$\sigma_N$.  Define a map $T_N: X_{N-1}\times B \to X_N$ as follows:
\begin{equation}\label{factor}
T_N(\vec y,v) = 
\left(\alpha(v) y_1 - {1\over \sqrt{N^2-N}}v, \dots,\alpha(v) y_{N-1} -
{1\over \sqrt{N^2-N}}v,  \sqrt{N-1\over N}v\right)\ ,
\end{equation}
where
$$\alpha^2(v) = 1 - |v|^2\ .$$

This map induces coordinates $(\vec y,v)$ on $X_N$, and in terms of these coordinates, one has
the integral factorization  formula 
$$\int_{X_N}\phi(\vec v){\rm d}\sigma_N = 
{|S^{3N-7}|\over|S^{3N-4}|}\int_{B}\left[\int_{X_{N-1}}\phi(T_N(\vec y,v)) {\rm d}\sigma_{N-1}\right]
(1 -|v|^2)^{(3N-8)/2} {\rm d}v\ .$$

It follows from this and the definition of $P_N$ that
$$P_N\phi(\vec v) =  f\circ \pi_N(\vec v)$$
where
$$f(v) = \int_{X_{N-1}}\phi(T_N(\vec y,v)) {\rm d}\sigma_{N-1}\ .$$

For $j< N$, one has analogous formulas for $T_j$ and $P_j$, except  the roles of $v_N$ and $v_j$ are interchanged.

Next,
  we  make the definition for $Q$ that is analogous to (\ref{pgap}) for $P$:
Define $\lambda_N$ by 
\begin{equation}\label{qgap}
\lambda_N = \sup\left\{\int_{X_N}\phi(\vec v) Q \phi(\vec v){\rm d}\sigma_N \ \bigg|\  
\int_{X_N}\phi^2(\vec v){\rm d}\sigma_N = 1\ ,\ \int_{X_N}\phi(\vec v){\rm d}\sigma_N = 0\ \right\}\ .
\end{equation}

With this explicit formula in hand, and the definitions of $\mu_N$ and $\lambda_N$, we come 
to the fundamental fact relating $P$ and $Q$:

\medskip
\begin{lm}\label{fund}
For any square integrable function $\phi$ on $X_N$ that is orthogonal to the constants,
\begin{equation}\label{rec}
\langle \phi, Q\phi\rangle \le \lambda_{N-1}\|\phi\|_2^2 + (1- \lambda_{N-1})
\langle \phi, P\phi\rangle\ ,
\end{equation}
where $\langle \cdot,\cdot\rangle$ denotes the inner product on $L^2(X_N,\sigma_N)$.
\end{lm}
\medskip

\noindent{\bf Proof:} To  bound  $\langle \phi, Q\phi\rangle$ in 
terms of $\lambda_{N-1}$,  define for $1\le k \le N$,
the operator $Q^{(k)}$ by
$$Q^{(k)}\phi(\vec v) =\left(
\begin{array}{c}
N-1\\ 2\end{array} \right)^{-1}\sum_{i<j,i\ne k,j\ne k}\int_{S^2}\phi(R_{i,j,\sigma}(\vec v)){\rm d}\sigma\ .$$
That is, we leave out collisions involving the $k$th particle, and average over the rest.
Clearly,
$$Q = {1\over N}\sum_{k=1}^N Q^{(k)}\ .$$
Therefore, for any  $\phi$  in $L^2(X_N,\sigma_N)$,
$$\langle \phi, Q\phi\rangle =  {1\over N}\sum_{k=1}^N 
\langle \phi, Q^{(k)}\phi\rangle\ .$$
Using the coordinates $(\vec y,v)$  induced by the map $T_k:X_{N-1}\times B \to X_N$,
it is easy to see that for $i\ne k,j\ne k$,  $R_{i,j,\sigma}$ acts only on the $\vec y$ variable.
That is, for such $i$ and $j$,
$$ R_{i,j,\sigma}(T_k(\vec y,v)) = T_k(R_{i,j,\sigma}(\vec y),v)\ .$$
Thus, if we hold $v$ fixed as a parameter,  we can think of $(Q^{(k)}\phi)(T_k(\vec y,v))$
as resulting from applying the $N-1$ dimensional version of $Q$ to $\phi$ with $v_k$ held fixed.

To estimate $\lambda_N$, we need estimate $\langle \phi, Q\phi\rangle$ when $\phi$ is orthogonal to the constants. When $\phi$ is  orthogonal to the constants, and we fix $v$, the function
$$\vec y \mapsto \phi(T_k(\vec y,v))$$
is not, in general,  orthogonal to the constants on $X_{N-1}$. However, we can correct for that by
adding and subtracting $P_k\phi$. Therefore
\begin{eqnarray}\label{ind1}
\langle (\phi - P_k\phi), Q^{(k)}(\phi-P_k\phi)\rangle &\le& \lambda_{N-1}\|\phi - P_k\phi\|_2^2 \nonumber\\
&=& \lambda_{N-1}(\|\phi\|_2^2 - \|P_k\phi\|_2^2)\nonumber\\
 &=& \lambda_{N-1}(\|\phi\|_2^2 -  \langle \phi, P_k\phi\rangle) \ .\nonumber\\
 \end{eqnarray}
Then since  $Q^{(k)}P_k\phi = P_k\phi$ and since $P_k\phi$ is orthogonal to
$\phi - P_k\phi$,
\begin{eqnarray}\label{ind2}
\langle \phi, Q^{(k)}\phi\rangle &=& \langle ((\phi - P_k\phi)+P_k\phi)Q^{(k)} ((\phi - P_k\phi)+P_k\phi)\rangle\nonumber\\
&=&  \langle (\phi - P_k\phi), Q^{(k)}(\phi-P_k\phi)  +  \langle P_k\phi, P_k\phi\rangle\nonumber\\
&=&  \langle (\phi - P_k\phi), Q^{(k)}(\phi-P_k\phi)\rangle  +  \langle \phi, P_k\phi\rangle\nonumber\\
&\le&   \lambda_{N-1}(\|\phi\|_2^2 -  \langle \phi P_k\phi\rangle)  + 
 \langle \phi P_k\phi\rangle\nonumber\\
 \end{eqnarray}
Averaging over $k$, we have (\ref{rec}). \qed

\medskip

Lemma \ref{fund}  was used as follows in \cite{CCL2}: Any trial function $\phi$  for $\lambda_N$ is a valid trial function
for $\mu_N$, so that
\begin{equation}\label{lamrec}
\lambda_N \le \lambda_{N-1} + (1-\lambda_{N-1})\mu_N\ .
\end{equation}
Then since $\Delta_N = N(1 - \lambda_N)$,
we have
\begin{equation}\label{delrec}
\Delta_N \ge {N\over N-1}(1 - \mu_N)\Delta_{N-1}\ .
\end{equation}
Therefore, with $a_N = {\displaystyle {N\over N-1}(1 - \mu_N)}$, for all $N \ge 3$,
$$\Delta_N \ge \left(\prod_{j =3}^N a_j\right)\Delta_2\ .$$
Thus, one route to proving a lower bound on $\Delta_N$ is to prove an upper bound on $\mu_N$, and hence
an lower bound on $a_N$.  This route led to a sharp lower bound for $\Delta_N$ --- the exact value --- for the one dimension Kac model investigated in \cite{CCL1}.  However, it would not lead to a proof of Theorem \ref{exact}.  The reasons for this are worth pointing out before we proceed:

As we shall see below, the eigenspace of $P$ with the eigenvalue $\mu_N$ --- the gap eigenspace of $P$  --- is spanned by the functions
specified in (\ref{eig1}).  Granted this, and granted Theorem \ref{exact},  whenever  condition (\ref{d2cond}) is satisfied:
\medskip
$${\rm For\ }  (1-B_2) < (1-B_1)\ ,\quad  Q\phi = \lambda_N\phi\quad \Rightarrow \quad P\phi = \mu_N\phi\ ,$$
while
$${\rm For\ }  (1-B_1) < (1-B_2)\ ,\quad  Q\phi = \lambda_N\phi\quad \Rightarrow \quad P\phi \ne \mu_N\phi\ .$$
\medskip

  In the second case, $ (1-B_1) < (1-B_2)$, the mismatch between the gap eigenspaces for $Q$ and $P$ means that
equality cannot hold in (\ref{lamrec}), and hence the recursive relation (\ref{delrec}) cannot possibly yield exact results in this case.

Moreover, even in the first case, $(1-B_2) < (1-B_1)$, where there is a match between the gap eigenspaces of $Q$ and $P$, there {\it still}   will not be equality in  (\ref{lamrec}).  The reasons for this are more subtle: The inequality (\ref{lamrec}) comes from the key estimate
(\ref{ind2}). Considering (\ref{ind2}), one sees that equality will hold there if and only if 
$$Q^{(k)}(\phi - P_k\phi) = \lambda_{N-1}
(\phi - P_k\phi)$$ for each $k$, where $(\phi - P_k\phi)$ is regarded as a function on $X_{N-1}$ through the change of variables
$T_k:(X_{N-1},B) \to X_N$ that was introduced just before Lemma \ref{fund}.

However, if $\phi$ is in the gap eigenspace for $Q$ on $X_N$,  Theorem \ref{exact} tells us that it is a linear combination of the three functions  specified in (\ref{eig1}), all of which are homogeneous of degree $3$ in $v$.  Because of the translation in (\ref{factor}),
which is due to momentum conservation,  $(\phi - P_k\phi)$ is regarded as a function on $X_{N-1}$ will {\it not} be homogeneous of degree
$3$ --- it will contain lower order terms. Hence $(\phi - P_k\phi)$ will not be in the gap eigenspace for $Q^{(k)}$.

The main result of the next section provides a way to use more detailed spectral information about $P$ to sharpen the recursive estimate
so that we do obtain the exact results announced in Theorem \ref{exact}.

\medskip
\section{How to use more detailed spectral information on $P$ to determine the spectral gap of $Q$} \label{howto}
\medskip

The following  lemma is the key to using (\ref{rec}) to obtain sharp results  for the  model considered here.

\medskip
\begin{lm}\label{mo} For any $N\ge 3$, let $\mu^\star_N$ be a number  with 
$$\mu_N^\star < \mu_N$$ such that there are only finitely eigenvalues
of $P$ between $\mu_N^\star$ and $\mu_N$:
$$\mu_N^\star \le  \mu_N^{(m)}  < \cdots <   \mu_N^{(1)} < \mu_N \ .$$

Let $\mu_N^{(0)}$ denote $\mu_N$, and then 
for $j=0,\dots, m$, let $E_j$ denote the eigenspace of $P$ corresponding to
$ \mu_N^{(j)} $.  Let $V_j$ denote the smallest invariant subspace of $Q$ that
contains $E_j$. Let $\nu_j$ be the largest eigenvalue of $Q$ on $V_j$. 

Then either 
\begin{equation}\label{oneway}
\lambda_N = \max\{\nu_0,\dots,\nu_{m}\}\ ,
\end{equation}
or else
\begin{equation}\label{twoway}
\Delta_N \ge \frac{N}{N-1}(1- \mu_N^\star)\Delta_{N-1}
\end{equation}
If  $\mu_N^\star =  \mu_N^{(m)}$, then we have the same alternative except with strict inequality in (\ref{twoway}).

\end{lm}

\medskip
\noindent{\bf Proof:}  If $\lambda_N > \max\{\nu_0,\dots,\nu_{m}\}$, then in the variational principle for
$\lambda_N$, we need only consider functions $\phi$ that are orthogonal to the constants, and also
in the orthogonal complement of each of the $V_j$.  This means also that $\phi$ belongs to the
orthogonal complement of each of the $E_j$. But then
$$\langle \phi, P\phi\rangle \le   \mu_N^\star \|\phi\|_2^2\ .$$
Using this estimate in (\ref{rec}), we have (\ref{twoway}).   Moreover, if $\mu_N^\star = \mu_N^{(m)}$, 
then strict inequality must hold in the last inequality. \qed
\medskip

Lemma \ref{mo} gives us the dichotomy between (\ref{indl2}) and \eqref{indl5} that plays a key role in the strategy described in the introduction.
To put this strategy into effect, we must first carry out a more detailed investigation of the spectrum of $P$.  The main result of the next section
reduces the investigation of the spectrum of $P$  to the study of  simpler operator --- the {\it correlation operator}  $K$, which is a Markov operator on functions on the unit ball $B$ in $\R^3$.

\medskip
\section{The correlation operator $K$, and its relation to $P$} \label{KO}
\medskip

While $Q$ and $P$ are both operators on spaces of functions of a large number of variables,
the problem of computing the eigenvalues of $P$ reduces to the problem of computing the eigenvalues of an operator on functions on $B$, the unit ball  in $\R^3$: 

First, define  the measure $\nu_N$ on $B$ to be the ``push forward'' of $\sigma_N$ under the map $\pi_j$.
That is, for any continuous function $f$ on $B$,
$$\int_B f(v){\rm d}\nu_N = \int_{X_N} f(\pi_j(\vec v)){\rm d}\sigma_N\ .$$
By the permutation invariance of $\sigma_N$, this definition does not depend on the choice of $j$.
By direct calculation \cite{CCL2}, one finds that
\begin{equation}\label{nufor}
{\rm d}\nu_N(v) = {|S^{3N-7}|\over|S^{3N-4}|}(1 -|v|^2)^{(3N-8)/2} {\rm d}v\ .
\end{equation}

Now define the self 
 adjoint operator  operator $K$ on $L^2(B,{\rm d}\nu_N)$  through the following 
quadratic form:
\begin{equation}\label{kopdef1}
\langle f, K f\rangle _{L^2(\nu)} = 
\int_{X_N}f(\pi_1(\vec v))f(\pi_2(\vec v)){\rm d}\sigma_N
\end{equation}
for all $f$ in $L^2(B,{\rm d}\nu_N)$. Equivalently,
\begin{equation}\label{kopdef}
(Kf)\circ \pi_1 = P_1(f\circ \pi_2)\ .
\end{equation}
Note that by the permutation invariance of $\sigma_N$, one can replace the pair $(1,2)$ of indices by any other pair of distinct indices without affecting the operator $K$ defined by (\ref{kopdef}).
This is the {\it correlation operator}.

To see the relation between the spectra of $P$ and the spectra of $K$, suppose that $\phi$ is an eigenfunction of $P$ that is symmetric under permutation of the particle indices.  (These symmetric eigenfunctions are the ones that are significant in the physical application.) 
Then since any vector in the image of $P$
has the form $\sum_{j=1}^Nf_j\circ \pi_j$ for functions $f_1,\dots,f_N$ on $B$, we must have, for $\phi$ symmetric, 
\begin{equation}\label{sym1}
\phi  = \sum_{j=1}^N f\circ \pi_j\ .
\end{equation}

Now we ask: For which choices of $f$ will $\phi$ given by (\ref{sym1}) be an eigenfunction of $P$?
To answer this, note that by
by (\ref{kopdef}),
\begin{equation}\label{pf}
P_k\phi =   f\circ \pi_k + \sum_{j=1,j\ne k}^N P_k(f\circ \pi_j)\ .
\end{equation}

Therefore,  from (\ref{pf}) and the definition of $K$,
$P_k\phi =   f\circ \pi_k + (N-1)(Kf)\circ\pi_k$.
Thus, averaging over $k$,
\begin{equation}\label{symm2}
P\phi = {1\over N}\phi + {N-1\over N}\sum_{j=1}^N (Kf)\circ \pi_j\ .
\end{equation}
In the case $Kf = \kappa f$, this reduces to 
$$P\phi = \frac{1}{N}(1+(N-1)\kappa)\phi\ ,$$
and thus eigenfunctions of $K$ yield eigenfunctions of $P$. 
It turns out that all symmetric eigenfunctions arise in exactly this way, and that
all eigenfunctions, symmetric or not, arise in a similar way, specified in the next lemma. 

\medskip
\begin{lm}\label{ktpp}
Let $V$ be the orthogonal complement in $L^2(X_N,\sigma_N)$ of the kernel of $P$.  There is a complete orthonormal basis of $V$ consisting of eigenfunctions $\phi$ of $P$ of one of the two forms:
\medskip

\noindent{(i)} For some eigenfunction $f$ of $K$,   
${\displaystyle\phi = \sum_{k=1}^N f\circ \pi_k}$.
In this case, if $Kf =  \kappa f$, then $ P\phi = \mu \phi$ where
\begin{equation}\label{bcn30}
\mu = \frac{1}{N}\left( 1 + (N-1){\kappa}\right)\ .
\end{equation}

\noindent{(ii)} For some eigenfunction $f$ of $K$,  and some pair of indices $i< j$, 
${\displaystyle\phi = f\circ \pi_i - f\circ \pi_j}$.
In this case, if $Kf =  \kappa f$,  then $ P\phi = \mu \phi$ where
\begin{equation}\label{bcn31}
\mu = \frac{1 - \kappa}{N}\ .
\end{equation}
\end{lm}

\medskip
\noindent{\bf Proof:}   Suppose that $\phi$ is an eigenfunction of $P$ with non zero eigenvalue $\mu$,
and  $\phi$ is orthogonal to the constants. By the permutation invariance
we may assume that either $\phi$ is invariant under permutations, or that there is some pair permutation,
which we may as well take to be $\sigma_{1,2}$, such that $\phi\circ\sigma_{1,2} = -\phi$. 
We will treat these two cases separately.

First suppose that $\phi$ is symmetric.  We have already observed that in this case, the recipe $\phi = \sum_{j=1}^Nf\circ \pi_j$, with $f$ an eigenfunction of $K$,  yields symmetric eigenfunctions of $P$.  We now show that all symmetric eigenfunctions of $P$ on $V$ have this form.

First, simply because such a $\phi$ is in the image of $P$, and is symmetric,  seen that $\phi$ must have the form (\ref{sym1}).  It remains to show that $f$ must be an eigenfunction of $K$. 
Then by (\ref{symm2}), $\mu\phi = P\phi$ becomes
$$\mu \sum_{k=1}^N f\circ\pi_k = {1\over N}\sum_{k=1}^N \left( f + (N-1)Kf\right)\circ\pi_k \ .$$
Apply $P_1$ to both sides to obtain
$${1\over N}\left(\left[ f + (N-1)Kf\right] + (N-1)K\left[ f + (N-1)Kf\right]\right) =
\mu(f + (N-1)Kf)$$
which is
\begin{equation}\label{j28a}
{1\over N}\left( I + (N-1)K\right)^2 f = \mu(I + (N-1)K)f\ .
\end{equation}
Since $\mu \ne 0$,   $f$  is not in the null space of either
$I + (N-1)K$ or $(I + (N-1)K)^2$. It then follows from (\ref{j28a}) that
$${1\over N}\left( I + (N-1)K\right) f = \mu f\ .$$
Thus,  when $\phi$ is symmetric, there is an eigenfunction $f$ of $K$ with eigenvalue $\kappa$, such that  $\phi = \sum_{k=1}^N f\circ \pi_k$  and
$$\mu = {1\over N}\left(1 +(N-1) \kappa\right)\ .$$

We next consider the case in which 
$$\phi\circ\sigma_{1,2} = -\phi\ .$$
Note that
$$P_k(\phi\circ\sigma_{1,2}) = P_k\phi=0$$
whenever $k$ is different from both $1$ and $2$. It follows that
$${1\over N}\sum_{k=1}^NP_k\phi = {1\over N}\left(P_1\phi + P_2\phi\right)\ .$$
The right hand side is of the form $f(v_1) - f(v_2)$, and hence $\phi$ must have this
form if it is an eigenvector. Taking $\phi = f\circ\pi_1 - f\circ\pi_2$
we have
$${1\over N}\sum_{k=1}^NP_k\phi  = {1\over N}\left(
(f-Kf)\circ\pi_1 - (f-Kf)\circ\pi_2\right)\ .$$
Hence when $P\phi = \mu \phi$ and $\phi$ is antisymmetric as above, There is an eigenvalue $\kappa$ of $K$
such that
$$\mu = {1- \kappa \over N}$$
This proves the second part. 
\qed

\medskip

Lemma \ref{ktpp} reduces the computation of the spectrum of $P$ to the computaton of the spectrum of $K$. 
We undertake this in the next three sections.

\medskip
\section{Explicit form of the correlation operator $K$} \label{K}
\medskip

For any two functions $f$ and $g$ on $B$ that are square integrable with respect to $\nu_N$,
 consider the bilinear form 
 $\int_{X_N}f(\pi_1(\vec v))g(\pi_2(\vec v)){\rm d}\sigma_N$.  It is easily seen from (\ref{kopdef}) that
 $$\langle f, Kg\rangle  =   \int_{X_N}f(\pi_1(\vec v))g(\pi_2(\vec v)){\rm d}\sigma_N\ ,$$
 where here, $\langle \cdot ,\cdot \rangle$ is the inner product on $L^2(B,\nu_N)$.
 
 Computing the right hand side using the factorization formula (\ref{factor}), but for $T_1$ instead of $T_N$,
 one finds, for $N > 3$:
 $$Kg(v) = 
  {|S^{3N-10}|\over|S^{3N-7}|}\int_{B}g\left({\sqrt{N^2-2N}\over N-1}
\sqrt{1-|v|^2}y - {1\over N-1}v\right)  (1 -|y|^2)^{(3N-11)/2} {\rm d}y\ .$$
The explicit form of $K$ is slightly different for $N=3$. We can see this different form as a limiting case, if we
make the dimension a continuous fact. The following way of doing this will be convenient later on:

For $\alpha > -1$, define the constant $C_\alpha$ by 
$$C_\alpha = \left(\int_B (1 -|y|^2)^\alpha  {\rm d}y\right)^{-1}\ ,$$
so that for 
$$\alpha = {3N -8\over 2}\ ,$$
$${\rm d}\nu_N(v)  = C_\alpha  (1 -|y|^2)^\alpha  {\rm d}y\ ,$$
and then 
$$Kg(v) = C_{\alpha -3/2}\int_B 
g\left({\sqrt{N^2-2N}\over N-1}
\sqrt{1-|v|^2}y - {1\over N-1}v\right)  (1 -|y|^2)^{\alpha -3/2} {\rm d}y \ .$$

Now, as $N$ approaches $3$, $\alpha -3/2$ approaches $-1$. Then the measure 
$C_\alpha  (1 -|y|^2)^\alpha  {\rm d}y$ concentrates more and more on the boundary of the ball $B$, so that
in the limit, it becomes the uniform measure on $S^2$. Understood in this way, the formula
remains valid at $\alpha = 1/2$; i.e., at $N=3$. 

It is clear that $K$ is a self adjoint Markov operator on $L^2(B,\nu_N)$, and that 
$1$ is an eigenvalue of multiplicity one.  With more effort, there is much more that can be said;
the spectrum of $K$ can be completely determined.

\medskip
\section{The  spectrum of $K$ and ratios of Jacobi polynomials} \label{detspeck1}
\medskip

In  studying the spectrum of the correlation  operator,
it is in fact natural and useful to study a 
wider family of operators of this type. Fix any $\alpha > 1/2$, and any numbers $a$ and $b$ such that
$$a^2+ b^2 =1\ .$$
Then define the generalized correlation operator, still simply denoted by $K$, through
\begin{equation}\label{genkdef}
Kg(v) = C_{\alpha -3/2}\int_B 
g\left(a \sqrt{1-|v|^2}y + b v\right)  (1 -|y|^2)^{\alpha -3/2} {\rm d}y \ .
\end{equation}

Notice that as $v$ and $y$ range over $B$, the maximum of  
$|a \sqrt{1-|v|^2}y + b v|$ occurs when  $ay$ and  $bv$ are parallel. In that case, 
$$|a \sqrt{1-|v|^2}y + b v| = |a||y|\sqrt{1-|v|^2} + |b||v|  1\le (a^2+b^2)^{1/2}((1- |v|^2)|y|^2 + |v|^2)^{1/2} \le 1\ .$$
Thus, as  $v$ and $y$ range over $B$, so does 
\begin{equation}\label{change}
u(y,v) =   a \sqrt{1-|v|^2}y + b v\ ,
\end{equation}
and $g(a \sqrt{1-|v|^2}y + b v)$ is well defined for any function $g$ on $B$. Thus, $K$ is well defined.

Now when 
\begin{equation}\label{fin}
a =  {\sqrt{N^2-2N}\over N-1}\qquad{\rm and}\qquad b = -{1\over N-1}\ ,
\end{equation}
we know that $K$ is self adjoint because in that case it is defined in terms of a manifestly 
symmetric bilinear form.   We shall show here that $K$ is always self adjoint for  all
$a^2+b^2 =1$, and that the eigenvalues of $K$ are given by an explicit formula involving
ratios of Jacobi polynomials.

To explain this, we fix some terminology and notation. For any numbers $\alpha>-1$ and $\beta>-1$,
$P_n^{(\alpha,\beta)}$ denotes the $n$th degree polynomial in the sequence of orthogonal polynomials
on $[-1,1]$ for the measure
$$(1-x)^\alpha(1+x)^\beta{\rm d}x\ ,$$
and is referred to as the $n$th degree Jacobi polynomial for $(\alpha,\beta)$. As is well known, 
$\{P_n^{(\alpha,\beta)}\}_{n\ge 0}$ is a complete orthogonal basis for $L^2([-1,1], (1-x)^\alpha(1+x)^\beta{\rm d}x)$.

Of course, what we have said so far specifies  $P_n^{(\alpha,\beta)}$ only up to a multiplicative constant. One common normalization is given by  Rodrigues' formula
$$P_n^{(\alpha,\beta)}(x) = \frac{(-1)^n}{2^n n!}(1-x)^{-\alpha}(1+x)^{-\beta}
\frac{{\rm d}^n}{{\rm d}x^n}\left((1-x)^{\alpha+n}(1+x)^{\beta+n}\right)\ .$$
For this normalization,
\begin{equation}\label{onenorm}
P_n^{(\alpha,\beta)}(1) = 
\left(
\begin{array}{c}
n+\alpha\\ n\end{array} \right) \qquad{\rm and}\qquad
P_n^{(\alpha,\beta)}(-1) = 
\left(
\begin{array}{c}
n+\beta\\ n\end{array} \right)\ .
\end{equation}

\medskip
\begin{lm}\label{rformlem}  Fix any $\alpha > 1/2$, and any numbers $a$ and $b$ such that
$a^2+ b^2 =1$, and define $K$ through the formula (\ref{genkdef}). Then $K$ is a self adjoint Markov operator, and the spectrum of $K$ consists of eigenvalues $\kappa_{n,\ell}$
enumerated by non negative integers $n$ and $\ell$, and these eigenvalues are 
given by the explicit formula
\begin{equation}\label{rf1}
\kappa_{n,\ell} = 
{P_n^{(\alpha,\beta)}(-1+2b^2)\over P_n^{(\alpha,\beta)}(1)}
b^{\ell}
\end{equation}
where $\beta = \ell+1/2$, $\alpha$ is the parameter $\alpha$ entering into the definition of $K$.
\end{lm}

\noindent{\bf Proof:} 
To see that $K$ is self adjoint, we write it as a bilinear form, and change variables to reveal the symmetry.
The change of variable that we make is naturally $(y,v) \to (u,v)$ with $u(y,v)$ given by (\ref{change}).
From (\ref{change}), one computes $y = u - bv/(a\sqrt{1- |v|^2}$, so that
\begin{eqnarray}\label{goodid}
1 - |y|^2 &=& {a^2 - a^2|v|^2 - |u|^2 - b^2|v|^2 - 2bu\cdot v\over b^2(1-|v|^2)}\nonumber\\
&=&{b^2 - (|u|^2+|v|^2) - 2bu\cdot v \over a^2(1- |v|^2}\ .\nonumber\\
\end{eqnarray}
The Jacobian is easy to work out, and one finds that ${\rm d}u{\rm d}v = a^3(1-|v|^2)^{3/2}{\rm d}y{\rm d}v$,
so that
\begin{eqnarray}\label{goodid2}
&\phantom{.}&\int_B f(v) Kg(v)C_\alpha (1-|v|^2)^\alpha {\rm d}v = \nonumber\\
&\phantom{.}&\int_B\int_Bf(v)g(u)a^{-2\alpha}\left[a^2 - (|u|^2+|v|^2) - 2bu\cdot v\right]_+^{\alpha-3/2} C_{\alpha-3/2}{\rm d}u{\rm d}v\ .\nonumber\\
\end{eqnarray}
This shows that the operator $K$ is self adjoint on $L^2(B, C_\alpha(1-|v|^2)^\alpha)$ for all $\alpha \ge 1/2$,
and all $a$ and $b$ with $a^2 + b^2 =1$. 
 
Our next goal is to prove
the eigenvalue formula (\ref{rf1}). 
This shall  follow from several simple properties of $K$.
 
First, $K$ commutes with rotations in
$\R^3$. That is, if $R$ is a rotation on $\R^3$, it is evident that 
$$K(g\circ R) = (Kg)\circ R\ .$$
Hence we may restrict our search for eigenfunctions $g$ of $K$ to functions of the form
$$g(v) = h(|v|)|v|^\ell {\cal Y}_{\ell,m}(v/|v|)$$
for some function $h$ on $[0,\infty)$, and some spherical harmonic ${\cal Y}_{\ell,m}$.

Second, for each $n\ge 0$, $K$ preserves the space of polynomials of degree $n$. To see this notice that
any monomial in $\sqrt{1-|v|^2}y$ that is of odd degree is annihilated
when integrated against $(1 -|y|^2)^{\alpha -3/2} {\rm d}y$,
and any even monomial in $\sqrt{1-|v|^2}y$ is a polynomial in $v$.

Combining these two observations, we see that $K$ has a complete basis of eigenfunctions of the form
$$g_{n,\ell,m}(v) = h_{n,\ell}(|v|^2)|v|^\ell {\cal Y}_{\ell,m}(v/|v|)$$
where $h_{n,\ell}$ is a polynomial of degree $n$. 

To determine these polynomials, we use the fact that $K$ is self adjoint, so that the eigenfunctions  $g_{n,\ell,m}$ can be taken to
be orthogonal.  In particular, for  
any two  distinct positive integers
$n$ and $p$, the eigenfunctions $g_{n,\ell,m}$ and $g_{p,\ell,m}$ are orthogonal in $L^2(B, C_\alpha(1-|v|^2)^\alpha)$. 
Hence for
each $\ell$, and for
$n \ne p$,
$$\int_{|v|\le 1} h_{n,\ell}(|v|^2)h_{p,\ell}(|v|^2)(1 - |v|^2)^{\alpha}|v|^{2\ell}{\rm d}v = 0\ .$$
Taking $r = |v|^2$ as a new variable, we have
$$\int_0^1 h_{n,\ell}(r)h_{p,\ell}(r)(1 - r)^{\alpha}r^{\ell+1/2}{\rm d}r = 0\ .$$
This is the orthogonality relation for a family of Jacobi polynomials in one standard form, and this identifies
the polynomials $ h_{n,\ell}$. A more common standard form, and one that is used in the sources to which we shall refer,
is obtained by the change of variable $t = 2r-1$, so that the variable $t$ ranges over the interval $[-1,1]$.
Then for $\alpha,\beta >-1$, $P_n^{(\alpha,\beta)}(t)$ is the $n$th degree orthogonal  polynomial for the weight
$(1-t)^\alpha(1+t)^\beta$.
With the variables $t$ and $|v|^2$ related as above; i.e.,
$$t = 2|v|^2 - 1\ ,$$
$$h_{n,\ell}(|v|^2) = P_n^{(\alpha,\beta)}(t)$$
for 
$$\beta = \ell+ {1\over 2}\ .$$

Now that we have all of the eigenfunctions determined, a further observation gives us a simple formula for the eigenvalues.
Consider any eigenfunction $g$ with eigenvalue $\kappa$, so  that $Kg(v) = \kappa g(v)$. Let ${\hat e}$ be any unit vector in $R^3$. Then since $g$ is a polynomial and hence
continuous,
\begin{eqnarray}
\lim_{t\to 1}Kg(t{\hat e}) &=& 
\lim_{t\to 1}\int_{B}g\left(a\sqrt{1-t^2}y +
b t{\hat e}\right) C_{\alpha -3/2}(1- |y|)^{\alpha -3/2}\d y\nonumber\\
&=& g\left(b{\hat e}\right)\ ,\nonumber\\
\end{eqnarray}
since $K1 = 1$. Combining this with $Kg(v) = \kappa g(v)$, we have
$$ g\left(b{\hat e}\right)  = \kappa g({\hat e})\ .$$

Now consider any eigenfunction $g_{n,\ell,m}$ of the form given above, and let $\kappa_{n,\ell}$
be the corresponding eigenvalue, which will not depend on $m$. Then taking any ${\hat e}$ so that ${\cal Y}_{\ell,m}(\hat
e)\ne 0$, we have  that
\begin{equation}\label{bcnra}
\kappa_{n,\ell} = {h_{n,\ell}(b^2)\over h_{n,\ell}(1)}b^\ell\ .
\end{equation}
Changing variables  as above to express this as a ratio of Jacobi polynomials, we finally have
proved (\ref{rf1}).  \qed

One might expect the largest eigenvalues of $K$ to correspond to eigenfunctions that are polynomials of
low degree. After all, in a system of orthogonal polynomials, those with high degree will have many changes of sign, and one might expect considerable cancelation when applying an averaging operator,
such as $K$, to them.  Therefore, let us compute the $\kappa_{n,\ell}$ for low values of $n$ and $\ell$.
We find from (\ref{rf1}), using the value $b = -1/(N-1)$ from (\ref{fin}),  that 
\begin{equation}\label{neg}
\kappa_{0,1} = \kappa_{1,0} = \frac{-1}{N-1}\ ,
\end{equation}
 so that $\kappa_{n,\ell}$ is negative for
$n+\ell = 1$. For $n+\ell = 2$, we find from (\ref{rf1}),

\begin{eqnarray}\label{lamvals}
\kappa_{1,1}(N)&=&\frac{5N-3}{3(N-1)^3}\nonumber\\
\kappa_{2,0}(N)&=&\frac{(N-3)(15N^2-15N+4)}{3(3N-4)(N-1)^4}\nonumber\\
\kappa_{0,2}(N)&=&\frac1{(N-1)^2}\ .\nonumber\\
\end{eqnarray}

Evidently, for large $N$,  
$$\kappa_{0,2}(N) = \frac{1}{N^2} + {\cal O}\left(\frac{1}{N^3}\right)\ ,$$ while
$$\kappa_{1,1}(N) = 
\frac{5}{3N^2} + {\cal O}\left(\frac{1}{N^3}\right)\qquad{\rm and}\qquad
\kappa_{0,2}(N) = 
\frac{5}{3N^2} + {\cal O}\left(\frac{1}{N^3}\right)\ .$$
Thus, one might expect that at least for large values of $N$, $1$, $\kappa_{1,1}(N)$, $\kappa_{2,0}(N)$ and 
$\kappa_{0,2}(N)$
are the four largest eigenvalues of $K$, and that $\kappa_{0,1} = \kappa_{1,0}$ is the most negative, with all other eigenvalues of $K$  lying  strictly between these. 

We shall show in the next section that this is indeed the case for all $N\ge 4$,
and that $1$ and $\kappa_{1,1}$  are the two largest eigenvalues of $K$ for all $N\ge 3$. 

When we use Lemma \ref{ktpp} to convert this to spectral information on $P$, we find that
$\kappa_{0,1}$, $\kappa_{1,0}$ and $\kappa_{0,2}$ all correspond to the same eigenvalues of $P$,
namely
$$\frac{1}{N} \left(1 + \frac{1}{N-1}\right) =  \frac{1}{N} \left(1 + (N-1)\frac{1}{(N-1)^2}\right)  = \frac{1}{N-1}\ .$$ This is the eigenvalue of $P$ that shall play the role of $\mu_N^{(m)}$ in our application of Lemma \ref{mo}. 

Let us conclude this section by recording a number of useful calculations that can be made using 
(\ref{rf1}).

For $N=3$, we have 
\begin{equation}\label{bcn20}
\kappa_{1,1}(3) = \frac{1}{2}  >  \kappa_{2,2}(3) = \frac{13}{40} >  \kappa_{0,2}(3) = \frac{1}{4} >  \kappa_{2,0}(3) = 0\ .
\end{equation}

For $N=4$, we have 
\begin{equation}\label{bcn21}
\kappa_{1,1}(4) = \frac{17}{81} >   \kappa_{0,2}(4) = \frac{1}{9} > \kappa_{2,0}(4) = \frac{23}{243}\ .
\end{equation}

For $N=5$, we have 
\begin{equation}\label{bcn22}
\kappa_{1,1}(5) = \frac{11}{96} > \kappa_{2,0}(5) = \frac{19}{264}  >   \kappa_{0,2}(5) = \frac{1}{16} \ .
\end{equation}

In each case, the second largest eigenvalue after $1$, among the ones listed,  is $\kappa_{1,1}$.
In the next section we shall see that the list is not misleading:  $\kappa_{1,1}$ is the gap eigenvalue.
However, note that the third largest eigenvalue comes from
different values of $n$ and $\ell$ for each of $N = 3$, $N =4$ and $N=5$. As we shall see, things do settle down for $N \ge 5$;
 the third largest eigenvalue does turn out to be $\kappa_{2,0}$ in all such cases.

\begin{lm}\label{order}
For all $N\ge 5$, $\kappa_{1,1}(N) > \kappa_{2,0}(N) > \kappa_{0,2}(N)$.
\end{lm}

\noindent{\bf Proof:}  From (\ref{lamvals}),
$$\kappa_{2,0}(N)-\kappa_{0,2}(N)=\frac{2N(3N^2-15N+8)}
{3(3N-4)(N-1)^4}\ .$$ 
A simple calculation shows that  the roots of the  polynomial in the numerator
are less than $5$, so that $\kappa_{2,0}(N)>\kappa_{0,2}(N)$ for $N\ge 5$.
A similar argument applied to $\kappa_{1,1}(N)-\kappa_{2,0}(N)$ yields the
conclusion of the lemma. \qed

\medskip

Our goal in the next section is to show that for all $N\ge 4$, there are no eigenvalues 
$\kappa_{n,\ell}$ with $n+\ell > 2$ that are larger that the ones listed above, and that for $N=3$,
the three largest eigenvalues are
$1= \kappa_{0,0}>1/2 =\kappa_{1,1}> 13/40 = \kappa_{2,2}$.  
However, since there is no simple monotonicity in $n+\ell$, this shall require some detailed estimate on ratios of Jacobi polynomials.

We shall also need to know that in all case $\kappa_{0,1} = \kappa_{1,0} = -1/(N-1)$ is the most negative eigenvalue.
This will tell us the four largest eigenvalues of $P$ for $N\ge 4$, and the three largest for $n=3$, and this shall turn out to be enough to prove the main result, Theorem \ref{exact}.

Finally, the value of $\kappa_{2,2}(N)$ will play an important role in the proof of Theorem \ref{exact2}, and so we record  the
expression here:

\begin{equation}\label{kap22val}
\kappa_{2,2}(N) = \frac{21 N^3-60 N^2+27 N -4 }{(3 N-4)(N-1)^6}\ .
\end{equation}

\medskip
\section{The determination of the spectrum of $K$} \label{detspeck2}
\medskip

The main result in this section is the following theorem:

\medskip

\begin{thm}\label{mono}
For $N\ge 5$ and all $n$ and $\ell$ with $n+\ell >2$, 
\begin{equation}\label{nge5}
-\frac{1}{N-1} \le \kappa_{n,\ell}(N)<\kappa_{0,2}(N)\ .\end{equation}
For $N= 4$ and all $n$ and $\ell$ with $n+\ell >2$
\begin{equation}\label{nge4}
-\frac{1}{N-1} \le \kappa_{n,\ell}(4)<\kappa_{2,0}(4)\ .\end{equation}
For $N=3$ and  all $n$ and $\ell$ with $n+\ell > 0$, except for $n=1,\ell =1$, 
\begin{equation}\label{nge3}
-\frac{1}{N-1} \le \kappa_{n,\ell}(3)\le\kappa_{2,2}(3)= \frac{13}{40} \ .\end{equation}
\end{thm}

\medskip

We present the proof at the end of this section after a number of preparatory lemmas.
These lemmas rest on two deep results about Jacobi polynomials. One is a formula
due to Koornwinder \cite{Kor}  (see also \cite{A}, pp. 31 {\it ff.})that was already applied in \cite{CCL2}:

For all $-1 \le x \le 1$, all $n$ and all $\alpha>\beta$,
\begin{equation}\label{Koor}{J_n^{(\alpha,\beta)}(x)\over J_n^{(\alpha,\beta)}(1)} = 
\int_0^{\pi}\int_0^1\left[{1+x-(1-x)r^2\over 2} + i\sqrt{1-x^2}r\cos(\theta)\right]^n
{\rm d}m_{\alpha,\beta}(r,\theta)
\end{equation}
where
$$m_{\alpha,\beta}(r,\theta) = c_{\alpha,\beta}(1-r^2)^{\alpha-\beta-1}
r^{2\beta+1}\left(\sin\theta\right)^{2\beta}{\rm d}r{\rm d}\theta\ ,$$
and $c_{\alpha,\beta}$ is a normalizing constant that 
makes ${\rm d}m_{\alpha,\beta}$ a probability measure.

Koornwinder's bound is very useful for obtaining uniform control in $n$ for given $\ell$ and $N$.
But since in Lemma \ref{rformlem}, 
\begin{equation}\label{crit}
\alpha = \frac{3N-8}{2}\quad{\rm and}\qquad \beta = \ell + \frac{1}{2}\ ,
\end{equation}
we can only apply (\ref{Koor}) when
\begin{equation}\label{lstardef}
\ell < \ell^\star = \frac{3N - 9}{2}\ \end{equation}
As in \cite{CCL2}, one may use this formula to show:

\begin{lm}\label{kblem}
For all $\ell$ with $2 \le \ell < \ell^\star$, and all $n>0$ and $N\ge 3$,
$$|\kappa_{n,\ell}(N)| < \frac{1}{(N-1)^2} = \kappa_{0,2}(N)\ .$$
\end{lm}

\medskip

Note that while this lemma does not address the case $n=0$, this is not a problem: we have the explicit formula
\begin{equation}\label{nez}
\kappa_{0,\ell} = \left(\frac{-1}{N-1}\right)^\ell\ .
\end{equation}

To handle large values of $\ell$, we need another deep result, which is
 a uniform
pointwise bound on the {\it orthonormal} Jacobi polynomials that
was obtained by
Nevai, Erdelyi, and Magnus. \cite{NEM}:  Let $p^{\alpha,\beta}_n$ be the orthonormal
Jacobi polynomial of degree $n$ with positive leading coefficient for the weight
$w(x)=(1-x)^{\alpha}(1+x)^{\beta}$.   It was shown in \cite{NEM} that for all
 $\alpha\ge -1/2$ and  $\beta\ge
  -1/2$ and all non negative integers $n$, 
\begin{equation}\label{nembnd}
{\rm max}_{x\in[-1,1]}\sqrt{1-x^2}w(x)p^{\alpha,\beta}_n(x)^2\le\frac{2e(2+\sqrt{\alpha^2+\beta^2})}{\pi},
\end{equation}

Of course, we could use the orthonormal Jacobi polynomials in the ratio formula (\ref{rf1}),
since any normalization factor would cancel out in the ratio. However, the exact formula
(\ref{onenorm})
for the denominator in (\ref{rf1}) is simplest in the other normalization. Hence we need
the relation between $p_n^{\alpha,\beta}$ and $P_n^{\alpha,\beta}$, which  is given by
$p_n^{\alpha,\beta}=l_n P_n^{\alpha,\beta}$ where 
$$
l_n=\left(\frac{2n+\alpha+\beta+1}{2^{\alpha+\beta+1}}\frac{\Gamma(n+1)\Gamma(n+\alpha+\beta+1)}{\Gamma(n+\alpha+1)\Gamma(n+\beta+1)}\right)^{1/2}.
$$
Therefore
\begin{equation}\label{nembnd2}
\frac{P_n^{\alpha,\beta}(x)^2}{P_n^{\alpha,\beta}(1)^2}\le
\frac{1}{l_n^2}\frac{2e\Gamma(n+1)^2\Gamma(\alpha+1)^2(2+\sqrt{\alpha^2+\beta^2})}{\sqrt{1-x^2}w(x)\pi\Gamma(n+\alpha+1)^2}.
\end{equation}

At this point it is perhaps worth noting  that since the spectrum of $K$ lies in $[-1,1]$, any upper bound on its eigenvalues by a number larger than one is vacuous. This implies that for certain regions the identity (\ref{bcnra}) will provide a stronger bound than (8.9). We shall return to this point at the end of the paper.

Substituting ${\displaystyle x=-1+\frac{2}{(N-1)^2}}$, ${\displaystyle \beta=\ell+\frac{1}{2}}$, $
{\displaystyle \alpha=\frac{3}{2}N-4}$ in (\ref{nembnd}), and  then multiplying by ${\displaystyle 
\frac{1}{(N-1)^{2l}}}$ yields
\begin{equation}\label{keycomp}
\kappa_{n,\ell}(N)\le\tilde\kappa_{n,\ell}(N)\ ,
\end{equation}
where
\begin{equation}\label{big1}
\tilde\kappa^2_{n,\ell} (N)  = \frac{2e}\pi g_1(n,\ell,N)g_2(N) g_3(n,N) g_4(n,\ell,N)
\end{equation}
where
\begin{eqnarray}\label{big}
g_1(n,\ell,N) &=&
\left(\frac{4+\sqrt{9N^2-48N+65+4\ell^2+4\ell}}
{3N+4n+2\ell-5}\right) \nonumber\\
g_2(N) &=&
\left(\frac{(N-1)^2}{N(N-2)}\right)^{(3N-7)/2}\nonumber\\
g_3(n,N) &=&
\frac{\Gamma(n+1)\Gamma\left(\frac32 N-3\right)}
{\Gamma \left(n+\frac32 N-3\right)}\nonumber\\
g_4(n+\ell,N) &=&
\frac{(N-1)^2\Gamma \left(n+\ell+\frac32\right)
\Gamma\left(\frac32 N-3\right)}{\Gamma \left(n+\ell + \frac32 N
-\frac52\right)}\nonumber\\
\end{eqnarray}

Our goal now is to extract a reasonably tight  upper bound for  $\tilde\kappa_{n,\ell}(N)$ with 
as as much monotonicity in $n$, $\ell$ and $N$ as possible.  The next lemmas address this goal.
\medskip

\begin{lm}\label{lem1} For $\ell\ge 0$, $N\ge 3$, and $n\ge 0$
\begin{equation}\label{g1bnd}
g_1(n,\ell,N) \le \left(\frac{4}
{3N+4n+2\ell-5} + 1\right)\ ,
\end{equation}
where the right hand side is clearly decreasing in $n$, $\ell$ and $N$.
\end{lm}

\noindent
\noindent{\bf Proof:}\ 
Note that for $n\ge 0$, $\ell \ge 0$ and $N\ge 3$,
\begin{equation} 
\frac{\sqrt{9N^2-48N+65+4\ell^2+4\ell}}{3N+4n+2\ell-5}\le 1
\end{equation}
since  then
\begin{align}
&(3N+4n+2\ell-5)^2-(9N^2-48N+65+4\ell^2+4\ell)\nonumber\\
&\quad = (24N-40)n+(12N-24)\ell+16n^2+16n\ell + 18N-40>0\ .
\end{align}
\qed

\begin{lm}\label{lem2} For  $N\ge 4$, $g_2(N)$ is a decreasing function of $N$.
\end{lm}

\medskip

\noindent
\noindent{\bf Proof:}\ Let $h(x) = (1 - 1/x^2)^{2-3x/2}$, so that $g_2(N) = h(N-1)$. 
Computing the derivative of $\ln(h(x))$, one finds that it is negative for $x\ge 3$. \qed

\medskip

\begin{lm}\label{lem3} For $n\ge 0$ and $N\ge 3$, $g_3(n,N)$ 
is a decreasing function of $n$ and $N$.
\end{lm}

\medskip
\noindent
\noindent{\bf Proof:}
For $n$ a nonzero integer 
\begin{equation}
\frac{\Gamma (n+1)\Gamma \left(\frac32 N-3\right)}
{\Gamma \left(n+\frac32 N-3\right)}=
\frac n{n+\frac32 N-4}\; \frac{n-1}{n+\frac32 N-5}
\cdots \frac1{\frac32 N-3}\,.\end{equation}
Since each factor is less than 1 for $N\ge 3$ and is a decreasing
function of $N$ the assertion follows. \qed
\medskip

\begin{lm}\label{lem4} 
For $N\ge 3$, $g_4(n+\ell,N)$ is a decreasing function of $n+\ell$
with
\begin{equation}\label{tozer}
\lim_{n+\ell \to \infty}g_4(n+\ell,N) = 0\ .
\end{equation}

Moreover, for  $n+\ell  \ge \ell^\star = 3(N-3)/2$,
\begin{equation}\label{g4bnd}
g_4(n,\ell,N) \le \frac{(N-1)^2\Gamma\left(\frac32 N-3\right)^2}{\Gamma
  (3N-7)}\le f(N)
\end{equation}
where  
\begin{equation}\label{g4bnd2}
f(N) = \frac{(N-1)^2\sqrt{\pi}(\frac{3}{2}N-4)}{2^{3N-8}}\ .
\end{equation}
Finally, for $N\ge 5$, $(N-1)^4f(N)$ is a decreasing function of $N$.
\end{lm}

\medskip
\noindent{\bf Proof:}  Since
\begin{equation}
\frac{\Gamma \left(n+\ell + \frac52\right)}{\Gamma\left( n+\ell +\frac32
N-\frac 32\right)}\;\frac{\Gamma\left(n+\ell + \frac32 N-\frac52\right)}
{\Gamma \left(n+\ell + \frac32\right)} = 
\frac{\left(n+\ell + \frac32\right)}{n+ \ell + \frac32 N-\frac52}
<1
\end{equation}
for $N\ge 3$, it follows that 
for a fixed nonnegative integers $N$,
\begin{equation}
\frac{\Gamma \left(n+\ell+\frac32\right)}{\Gamma\left(
n+ \ell+ \frac32 N-\frac52\right)}
\end{equation}
is a decreasing function of $n+\ell$.
Hence,  for $n+\ell \ge 3(N-3)/2$,
$$\frac{\Gamma \left(n+\ell+\frac32\right)}{\Gamma\left(
n+ \ell+ \frac32 N-\frac52\right)} \le 
\frac{\Gamma \left(\frac{3}{2}N-3\right)}{\Gamma\left(
3N-7\right)}$$
This together with the definition of $g_4$  proves the first inequality in 
(\ref{g4bnd}). Use of the duplication formula for the $\Gamma$ function
yields
$$
\frac{\Gamma\left(\frac32 N-3\right)^2}{\Gamma
  (3N-7)}=\frac{\sqrt{\pi}\Gamma\left(\frac32
    N-3\right)}{2^{3N-8}\Gamma\left(\frac32
    N-\frac72\right)}=\frac{\sqrt{\pi}\left(\frac32
    N-4\right)\Gamma\left(\frac32
    N-4\right)}{2^{3N-8}\Gamma\left(\frac32
    N-\frac72\right)}<\frac{\sqrt{\pi}\left(\frac32
    N-4\right)}{2^{3N-8}}.
$$
This implies the second inequality in (\ref{g4bnd}). A check of the
logarithmic derivative of $(N-1)^4 f(N)$ shows it is negative for $N\le 5$.

\qed
\medskip

Now, combining the results in the last four lemmas, we have that for  $N\ge 3$ and $n+\ell \ge 
\ell^\star = 3(N-3)/2$,
\begin{equation}\label{kap}
\tilde \kappa^2_{n,\ell}(N) \le \hat\kappa^2_{n,l}(N)\le \kappa^2(N)
\end{equation}
where
\begin{equation}\label{hatl}
\hat\kappa_{n,l}^2(N) = \frac{2e}{\pi} \left(\frac{4}
{3N+4n+2\ell-5} + 1\right)g_2(N)g_3(n,N)g_4(n,l,N)\ ,
\end{equation}
and
\begin{equation}\label{kap2}
\kappa^2(N) = \frac{2e}{\pi} \left(\frac{4}
{6N-14} + 1\right)g_2(N)f(N)\ ,
\end{equation}
where $g_2$, $g_3$ and $f$ are given by (\ref{big}) and  (\ref{g4bnd2}).

We are now ready to prove the main theorem of this section:
\medskip

\noindent{\bf Proof of Theorem \ref{mono}}  First, we take care of large values of $N$. 
By Lemmas \ref{lem2} and \ref{lem4}, $(N-1)^4\kappa(N)$ is a decreasing functions of
$N$ for $N\ge 5$. Direct computation shows that at $N=12$, this quantity is less than one.
Hence for $N\ge 12$,  $\kappa(N) \le (N-1)^{-4} = \kappa^2_{0,2}$.  For $\ell \ge \ell^\star$,
so that (\ref{kap}) is satisfied, this proves (\ref{nge5}) for $N\ge 12$.  On the other hand, if
$2 \le \ell < \ell^\star$, we have this from Lemma \ref{kblem} or (\ref{nez}). Thus, in any case, 
(\ref{nge5}) is valid for $N\ge 12$.

For $4 \le N \le 11$, we again use  Lemma \ref{kblem} or (\ref{nez}) for  
$2 \le \ell < \ell^*$, and computation of $\hat \kappa_{n,\ell}$. By (\ref{tozer}),
for each such $N$
there is a finite value $k(N)$ so that we need only consider values of $n+\ell < k(N)$. 
Checking these cases, we obtain (\ref{nge5}) and (\ref{nge4}). 

We finally turn to $N=3$, which requires the greatest amount of computation.
First for $n=0$, we have from (\ref{nez}) that 
$$\kappa_{0,\ell} (3)=\left(\frac{-1}2\right)^\ell$$
so $\kappa_{0,1}(3)=-1/2$ and $|\kappa_{0,\ell}(3)|<1/3$ for $\ell \ge 2$.

The exact forms of the eigenvalues are simple enough to be useful for $n=1$ and $2$ as well.
We have:
\begin{align*}
\kappa_{1,\ell} (N)&=\frac{(-1)^{\ell+1}}3
\frac{[2\ell N+3(N-1)]}{(N-1)^{\ell+2}}\\
\noalign{\noindent and}
\kappa_{2,\ell}(N)&=\frac{(-1)^\ell ((4\ell^2+16\ell + 15)N^3-
(8\ell^2+44\ell + 60)N^2+(49+16\ell)N-12)}{3(3N-4)(N-1)^{\ell+4}}
\end{align*}

Specializing to $N=3$, 
$$\kappa_{1,\ell}(3)=(-1)^{\ell+1}\frac{\ell+1}{2^{\ell+1}}$$
so that  $|\kappa_{1,\ell}(3)|\le 3/8$ for $\ell \ge 2$.  Likewise, for $N=3$,
$$\kappa_{2,\ell}(3)=\frac\ell{20}
\frac{(7+3\ell)}{2^\ell}(-1)^\ell$$ which implies that
\begin{equation}
|\kappa_{2,\ell}(3)|\le|\kappa_{22}(3)|=\frac{13}{40}
\end{equation}

For higher values of $n$, we estimate $\kappa^2_{n,\ell}$ by means of 
$\hat \kappa^2_{n,\ell}$. Since $\ell^\star =0$ for $N=3$, we may use Lemma \ref{lem4}
for all $\ell$, and then by (\ref{tozer}), for each fixed $n$, there is a maximal value $\ell(n)$
that need to be considered, and a maximum value of $n$ that need to be considered. 
The following table gives the values
of $n,\ell$ and $\hat\kappa^2_{n,\ell}(3)$ when $\tilde \kappa^2_{n,\ell}(3)<
1/4$. The monotonicity of $\kappa^2_{n,\ell}(3)$ in $n$ and $l$ shows
that $\hat\kappa^2_{n,\ell}(3)\le\hat\kappa^2_{n_0,\ell_0}(3)$ for $n\ge n_0$
and $\ell\ge\ell_0$ where
$(n_0,\ell_0)$ is chosen from the table.
\qed

\bigskip
$$ \begin {array}{ccc} n&\ell&\hat\kappa^2_{n,\ell}\\
\noalign{\medskip}3&1253& 0.10562
\\\noalign{\medskip}4&989& 0.10562\\\noalign{\medskip}5&817& 0.10556
\\\noalign{\medskip}6&694& 0.10561\\\noalign{\medskip}7&604& 0.10555
\\\noalign{\medskip}8&533& 0.10560\\\noalign{\medskip}9&477& 0.10558
\\\noalign{\medskip}10&431& 0.10558\\\noalign{\medskip}11&393& 0.10555
\\\noalign{\medskip}12&360& 0.10561\\\noalign{\medskip}13&333& 0.10548
\\\noalign{\medskip}14&308& 0.10558\\\noalign{\medskip}15&287& 0.10554
\\\noalign{\medskip}16&268& 0.10556\\\noalign{\medskip}17&251& 0.10558
\\\noalign{\medskip}18&236& 0.10554\\
\noalign{\medskip}19&222& 0.10560 \\
\end{array}\quad
 \begin {array}{ccc} n&\ell&\hat\kappa^2_{n,\ell}\\
\noalign{\medskip}20&210& 0.10547\\
\noalign{\medskip}21&198& 0.10559
\\\noalign{\medskip}22&188& 0.10546\\\noalign{\medskip}23&178& 0.10552
\\\noalign{\medskip}24&169& 0.10549\\\noalign{\medskip}25&161& 0.10537
\\\noalign{\medskip}26&153& 0.10540\\\noalign{\medskip}27&145& 0.10558
\\\noalign{\medskip}28&138& 0.10561\\\noalign{\medskip}29&132& 0.10542
\\\noalign{\medskip}30&126& 0.10534\\\noalign{\medskip}31&120& 0.10540
\\\noalign{\medskip}32&114& 0.10554\\\noalign{\medskip}33&109& 0.10543
\\\noalign{\medskip}34&104& 0.10540\\\noalign{\medskip}35&99& 0.10546
\\
\noalign{\medskip}36&94& 0.10562\\
\end{array}\quad
 \begin {array}{ccc} n&\ell&\hat\kappa^2_{n,\ell}\\
\noalign{\medskip}37&90& 0.10543\\
\noalign{\medskip}38&86& 0.10531\\
\noalign{\medskip}39&82& 0.10527
\\\noalign{\medskip}40&78& 0.10528\\\noalign{\medskip}41&74& 0.10537
\\\noalign{\medskip}42&70& 0.10551\\\noalign{\medskip}43&67& 0.10523
\\\noalign{\medskip}44&63& 0.10552\\\noalign{\medskip}45&60& 0.10534
\\\noalign{\medskip}46&57& 0.10521\\\noalign{\medskip}47&54& 0.10512
\\\noalign{\medskip}48&50& 0.10562\\\noalign{\medskip}49&48& 0.10508
\\\noalign{\medskip}50&45& 0.10512\\\noalign{\medskip}51&42& 0.10521
\\\noalign{\medskip}52&39& 0.10535\\
\noalign{\medskip}53&36& 0.10552 \\
\end{array}\quad
 \begin {array}{ccc} n&\ell&\hat\kappa^2_{n,\ell}\\
\noalign{\medskip}54&34& 0.10514\\
\noalign{\medskip}55&31& 0.10538 \\\noalign{\medskip}56&29& 0.10506\\
\noalign{\medskip}57&26& 0.10538
\\\noalign{\medskip}58&24& 0.10511\\\noalign{\medskip}59&21& 0.10551
\\\noalign{\medskip}60&19& 0.10529\\\noalign{\medskip}61&17& 0.10509
\\\noalign{\medskip}62&14& 0.10560\\\noalign{\medskip}63&12& 0.10545
\\\noalign{\medskip}64&10& 0.10534\\\noalign{\medskip}65&8& 0.10523
\\\noalign{\medskip}66&6& 0.10514\\\noalign{\medskip}67&4& 0.10509
\\\noalign{\medskip}68&2& 0.10506\\\noalign{\medskip}69&0& 0.10503
\\\noalign{\medskip} &&\\
\end {array} $$

\bigskip\noindent
The remaining values can be computed from the exact formula
for $\kappa_{n,\ell }(3)$ from (\ref{rformlem}),  and the results are all consistent with (\ref{nge3}).
\qed

\medskip

\medskip
\section{The determination of the spectrum of $P$}
\medskip

For given values of $N$, $n$ and $\ell$, 
let $\mu_{n,\ell}(N)$ be the eigenvalue of $P$ corresponding to the eigenvalue $\kappa_{n,\ell}(N)$ of $K$ through  Theorem \ref{mono},
where we use (\ref{bcn30}) if   $\kappa_{n,\ell}(N)>0$, and  use (\ref{bcn31}) if   $\kappa_{n,\ell}(N)<0$.  (This is the relevant choice, as we are concerned with the largest eigenvalues of $P$.)

Consulting the calculations in (\ref{neg}) for $n+\ell =1$, and in (\ref{bcn20}),  (\ref{bcn21}) and (\ref{bcn22}) for $n+\ell=2$, and
finally  the bounds in Theorem \ref{mono} for $n+\ell > 2$, we see that for all $N\ge 3$, the largest eigenvalues of $K$
is $\kappa_{1,1}$, and the least (most negative) is $\kappa_{0,1}= \kappa_{1,0}$.   Thus, turning to Lemma \ref{ktpp}, and using the positve eigenvalue in (\ref{bcn30}), and the negative one in (\ref{bcn31}), we see that the positive one yields the greater value for each $N$.
Thus, the gap eigenvalue of $P$, $\mu_N$, is given by 
\begin{equation}\label{gapval}
\mu_N = \mu_{1,1}(N) = \frac{3N-1}{3(N-1)^2}\ .
\end{equation}

Use of this result in (\ref{delrec}) would yield a strictly positive lower bound on $\Delta_N$, uniform in $N$, but, as we have said
above, it would not yield the exact lower bound.  To obtain this, we now carry out the strategy outlined in the introduction.

First, we  combine Lemma \ref{ktpp} and Theorem \ref{mono} to produce the  information necessary for the application of Lemma \ref{mo}.    We must now make a choice of  the thresholds $\mu_N^\star$ that appear in  Lemma \ref{mo}.
The choice we shall  make  is based on trial function computations with $Q$ that suggest that the gap eigenfunctions
are the ones specified in Theorem \ref{exact}.  

Notice that in Theorem \ref{exact}, the formula given for $\Delta_{N}$ is of the form ${\displaystyle C\frac{N}{N-1}}$ for some constant $C$.
This value can be guessed by computing the eigenvalues of $Q$ on the invariant subspace of polynomials of degree $4$
or less in the $v_j$.   If we are to prove this guess correct using (\ref{twoway}) of Lemma \ref{mo}, we require a value of $\mu_N^\star$
such that
\begin{equation}\label{bcn33}
\frac {N}{N-1}(1 - \mu_N^\star)\frac{N-1}{N-2} \ge   \frac {N}{N-1}\ ,
\end{equation}
at least for $N\ge 4$.  (The guess is valid only for  $N-1\ge 3$. For $N-1 =2$, there is a different value of $\Delta_2$
which has been determined already in Section 2.)

The largest value of $\mu_N^\star$ that will satisfy (\ref{bcn33}) is
\begin{equation}\label{bcn34}
\mu_N^\star = \frac{1}{N-1}\qquad{\rm for}\qquad N\ge 4\ .
\end{equation}
This turns out to be an eigenvalue of $P$: Indeed, we have found in (\ref{lamvals}) that $\kappa_{0,2} = 1/(N-1)^2$. 
Furthermore, we have found in (\ref{neg}) that   $\kappa_{0,1} = \kappa_{1,0} = -1/(N-1)$.
Using first of these results  in (\ref{bcn30}) of Lemma \ref{ktpp}, and the second in (\ref{bcn31}) we find
$$\mu_{0,2} = \mu_{1,0}= \mu_{0,1}  =   \frac{1}{N-1}\ .$$

For $N=3$ we need to make a different choice, as the spectrum of $Q$ is quite different for $N=2$ and for $N\ge 3$. 
The choice that shall work is $\mu_3^\star = \mu_{2,2}(3)$.   Since $\kappa_{2,2}(3) = 13/40$, we have from  Lemma  \ref{ktpp} that  ${\displaystyle
\mu_{2,2}(3) = \frac{1}{3}(1+2(13/40)) = \frac{11}{20}}$. 
Thus, 
\begin{equation}\label{bcn35}
\mu_3^\star = \frac{11}{20}\ .
\end{equation}

Now, to apply Lemma \ref{mo}, we need the eigenspaces of $P$ for the eigenvalues $\mu$ satisfying $1 > \mu > \mu^\star_N$. 
By Theorem \ref{mono} and (\ref{bcn21}),  for $N=3$ and $N=4$, there is just one such eigenvalue, namely $\mu_{1,1}(4)$, the gap eigenvalue, and for $N\ge 5$, there are two:  $\mu_{1,1}(N)$ and $\mu_{2,0}(N)$.

Let $E_{n,\ell}$ be the eigenspace of $P$ corresponding to the eigenvalue $\mu_{n,\ell}(N)$.  For all values of $n$ and $\ell$
with $n+\ell \le 2$, we have determined the corresponding eigenfunctions of $K$, and thus, through Lemma \ref{ktpp},
the corresponding eigenfunctions of $P$. Thus, we have the following explicit descriptions of the   $E_{n,\ell}$
for   $n+\ell \le 2$:

First,  for $n+\ell =1$, the eigenvalues of $K$ are negative, and so by Lemma \ref{ktpp}, the eigenfunctions are antisymmetric.
If we are only concerned with the spectrum of $Q$ on the subspace of symmetric functions (which is all that is of significance for
Kac's application to the Boltzmann equation), we can ignore these eigenspaces. However,  they turn out to be very simple. 
The $n=0,\ell =1$ eigenfuctions of $K$ are degree one spherical harmonics, and the  $n=1,\ell =0$ eigenfuctions of $K$ are degree one
Jacobi polynomials in $|v|^2$. Hence 
\begin{equation}\label{bcn40}
E_{0,1}\quad {\rm is  \ spanned \ by\ the\ functions}\quad    v_i^\alpha - v_j^\alpha\ , \alpha = 1,2,3 \quad{\rm and}\  i < j\ ,\ 
\end{equation}
while
\begin{equation}\label{bcn41}
E_{1,0}\quad {\rm is  \ spanned \ by\ the\ functions}\quad    |v_i|^2 - |v_j|^2\ , i < j\ ,\ 
\end{equation}

Next, for $n+\ell =2$, the eigenvalues of  $K$ are positive, and so by Lemma \ref{ktpp}, the eigenfunctions are symmetric.
The  $n=0,\ell =2$ eigenfuctions of $K$ are degree two spherical harmonics, and so have the form
$$f_{0,2}(v) = \sum_{\alpha,\beta =1}^3
{A_{\alpha,\beta}v^\alpha v^\beta}$$
for some traceless symmetric  $3\times 3$ matrix $A$. 
Hence, by  Lemma \ref{ktpp}, 
\begin{equation}\label{bcn42}
E_{0,2}\quad {\rm is  \ spanned \ by\ the\ functions}\quad   \sum_{j=1}^N f_{0,2}(v_j)  \ ,\ 
\end{equation}
with $f_{0,2}$ given as above.

For $n=1$, $\ell =1$, the eigenfunctions of $K$ are the product of a degree one spherical harmonic, and a degree one Jacobi polynomial
in $|v|^2$. When we sum over the particles, the constant term in the Jacobi polynomial drops out due to the  momentum constraint, and
we see that
\begin{equation}\label{bcn43}
E_{1,1}\quad {\rm is  \ spanned \ by\ the\ functions}\quad   \sum_{j=1}^N f_{1,1}(v_j) \ ,\ 
\end{equation}
where
$$f_{1,1}(v)  = |v|^2v^\alpha\quad \alpha =1,2,3\ .$$

Finally, for $n=2$, $\ell =0$, the eigenfunction of $K$ is a degree two Jacobi polynomial in $|v|^2$.  After summing on the particles, the linear term can be absorbed into the constant by the energy constraint, and so we see that 
\begin{equation}\label{bcn44}
E_{2,0}\quad {\rm is  \ spanned \ by\ the\ function}\quad   \sum_{j=1}^N f_{2,0}(v_j) \ ,\ 
\end{equation}
where
$$f_{2,0}(v)  = |v|^4  - \int_{B}|v|^4{\rm d}\nu_N\ .$$

We close this section with a lemma that we shall need to prove Theorem \ref{exact2}. There we shall need to know the next largest
eigenvalue of $P$ below $\max_{n+\ell \le 2}\mu_{n,\ell}(N)$.  One might guess that this occurs for some values of $n$ and $\ell$
with $n+\ell = 3$, but this is not the case: By (\ref{nge3}) of Theorem \ref{mono}, and (\ref{bcn20}),
we se that for $N=3$, the most negative eigenvalue of $K$ is $-1/2$, and by Lemma \ref{ktpp},
this corresponds to the eigenvalue $1/2$ of $P$.  On the other hand, the largest eigenvalue of $K$
appart from $\kappa_{1,1}(3)$ is $\kappa_{2,2}(3) = 13/40$. This corresponds to the eigenvalue
$11/20$ of $P$. Since $11/20>1/2$, we do indeed have that
$$\sup_{n+\ell>2}\mu_n,\ell(3) = \mu_{2,2}(3) = \frac{11}{20}\ .$$

It seems likely, on the basis of computations that we have made,  that in fact
\begin{equation}\label{fin4}
\sup_{n+\ell>2}\mu_n,\ell(N) = \mu_{2,2}(N)
\end{equation}
for all $N\ge 3$.  However, for the proof of Theorem \ref{exact2}, all that we require is:
\medskip

\begin{lm}\label{twotwo} 
For $N =3,4,5,6$ and $7$, (\ref{fin4}) is true.
\end{lm}

\medskip

\noindent{\bf Proof:}  Note that the case $N=3$ has already been proved in the remarks above.
To deal with the other cases, we proceed essentially as in the proof of Theorem \ref{mono}, using (\ref{hatl})  to reduce the numbe of cases to be checked to a finite number, and then checking these. We will therefore be brief in our remarks on the remaining cases. 

Perhaps the most important point to recall is that (\ref{hatl}) is valid for $n+\ell \ge  3(N-3)/2$.
since the right hand side evaluates to zero for $N=3$, we could use it without restriction. For
$N=7$ though, $3(N-3)/2$ evaluates to $6$, and so we may only use (\ref{hatl}) for
$n + \ell \ge 6$. So these cases must be checked be direct computation of the eigenvalues
using \ref{rformlem}), and then converting these to eigenvalues of $P$ using Lemma \ref{ktpp}.

Then, using  (\ref{hatl}) for $n+\ell>6$, one finds that
$$\kappa^2_{n,\ell}(7) <  \kappa^2_{2,2}(7)$$
unless $0\le n \le 6$ and $0\le \ell \le 27$. Computing the rest of the eigenvalues of $P$
in this $6$ by $27$ rectangle, we find that stated result is true for $N=7$. 

A similar analysis takes care of $N=4$, $N=5$, and $N=6$. 
\qed

\medskip

We shall not need to know the corresponding eigenfunctions in our application of Lemma \ref{twotwo}, since we will only be concerned with the eigenspaces of eigenvalues lying strictly above $\mu_{2,2}(N)$, and those have been determined already in this section.

\medskip
\section{The spectrum of  $Q$ on invariant subspaces containing eigenspaces of $P$}
\medskip

For each $n$ and $\ell$, let $V_{n,\ell}$ be the smallest invariant subspace of $Q$ constaining $E_{n,\ell}$.   As we shall see,
for $n+\ell \le 2$, $V_{n,\ell} = E_{n,\ell}$ except for $n=2$, $\ell = 0$, in which case $V_{2,0}$ is two dimensional, while $E_{1,0}$ is 
one dimensional. This is established in the next lemma, which also specifies the spectrum of $Q$ on these invariant subspaces.
 The eigenvalues of course depend on
the particular choice of $b$ in the definition of $Q$, but in a very simple way: The dependence on $b$ is only through
the quantities $(1-B_1)$ and  $(1-B_2)$, where $B_j$ is the $j$th moment of $b$,  as defined in (\ref{b2def}).

\begin{lm}\label{lemQ} 

Every nonzero function in $E_{0,1}$ and in $E_{1,0}$  is an eigenfunction of $Q$
with eigenvalue 
\begin{equation}\label{bcn61}
\lambda^Q_{0,1} =    \lambda^Q_{1,0}  =  1-(1-B_1)\frac{1}{N-1}\ ,
\end{equation}
so that $V_{0,1} = E_{0,1}$ and $V_{1,0} = E_{1,0}$.

Every nonzero function in $E_{1,1}$ is an eigenfunction of $Q$ with the eigenvalue 
\begin{equation}\label{oneoneq}
\lambda^Q_{1,1} = 1 - (1-B_2)\frac{1}{(N-1)}\ ,
\end{equation}
so that $V_{1,1} = E_{1,1}$.

Furthermore, every
nonzero function in $E_{0,2}$ is an eigenfunction of $Q$ with the eigenvalue 
\begin{equation}\label{otwoq}
\lambda^Q_{0,2} = 1 - (1-B_2)\frac{3}{2(N-1)}\ ,
\end{equation}
so that $V_{0,2} = E_{0,2}$.

Finally, while $V_{2,0}$ is larger than $E_{2,0}$, there are only two eigenvalues of 
$Q$ in $V_{2,0}$. These are
\begin{equation}\label{otwoq1}
1 - (1-B_2)\left(\frac{1}{N(N-1)}\left[(2N-1) \pm \sqrt{N^2-3N+1}\right]\right)\ .
\end{equation}

For all $N\ge 3$, the largest of these eigenvalues is $\lambda^Q_{1,1}$.
\end{lm}
\medskip

Before beginning the proof, we note that if $(1/2)b(x)\d x$ is a Dirac mass at $x=1$, the collisions are all trivial (zero scattering angle), and thus $Q = I$ in this case.  But also in this case $(1-B_1) = (1-B_2) = 0$, so all of the eigenvalues $\lambda_{n,\ell}^Q$
listed above  are $1$ --- as they must be for $Q = I$.

\medskip
\noindent{\bf Proof:} We begin with the last case, case $n=2$, $\ell =0$, which is the most involved.
Consider the function
\begin{equation}
\phi = \sum_{i=1}^N |v_i|^4 
\end{equation}
and note that $\phi - \int_{X_N}\phi{\rm d}\sigma_N$ spans $E_{2,0}$, as we have noted above.

One simple way to calculate $Q\phi$  is to take advantage of the permutation symmetry of $Q$: Define
the symmetrization operator ${\cal S}$ by
$${\cal S}f(v_1,\dots,v_N) = \frac{1}{N!}\sum_{\pi}f(v_{\pi(1)},\dots,v_{\pi(N)}) $$
where the sum runs over all permutations $\pi$ of $\{1,\dots,N\}$. Then it is easy to see that
$${\cal S}(|v_1|^4) = \frac{1}{N}\phi\ .$$
Thus, since ${\cal S}Q = Q{\cal S}$,
$$Q\phi = N{\cal S}Q(|v_1|^4) \ .$$
One now directly calculates   $Q(|v_1|^4)$ and then symmetrizes.
In carrying out the calculation, we make use of:

\begin{lm}\label{sigmaints} Let $c$ and $d$ be any two vectors in $\R^3$, and let $e$ be any unit vector in $\R^3$.
Then with $B_1$ and $B_2$ defined as in (\ref{b2def}),  we have the following identities:
$$\int_{S^2}(c\cdot\sigma) b(e\cdot\sigma)\d \sigma = (c\cdot e)B_1$$
and
$$\int_{S^2}(c\cdot\sigma) (d\cdot\sigma)  b(e\cdot\sigma)\d \sigma  = (c\cdot d)\frac{1-B_2}{2} + (e\cdot c)(e\cdot d)\frac{3B_2-1}{2}\ . $$
\end{lm}

\noindent{\bf Proof:} We choose coordinates in which
 $c$ and $e$ span the $x,z$ plane with $e = \left[
 \begin{array}{c}
0\\ 0\\1\end{array}\right]$ and $c = \left[ \begin{array}{c}  c^1 \\ 0\\ c^3 \end{array}\right]$.
Then with $\sigma = \left[\begin{array}{c} 
 \sin\theta\sin\varphi\\ 
 \cos\theta\sin\varphi\\ 
 \cos\theta\end{array}
 \right]$, the computations are easily accomplished. \qed.

Now to compute $Q\phi$, go back to the definition of $Q$ given in (\ref{qopdef}); and note first of all that with $\eta(\vec v) = |v_1|^4$, 
unless $i=1$,
$$\eta(R_{i,j,\sigma}(\vec v)) = \eta(\vec v)\ .$$
Hence
$$Q\eta(\vec v) = \left(1 - \frac{2}{N}\right)\eta(\vec v)
 + \frac{2}{N(N-1)}\sum_{j=2}^N\int_{S^2} \eta(R_{1,j,\sigma}(\vec v))b\left(\sigma\cdot \frac{v_i-v_j}{|v_i-v_j|}\right){\rm d}\sigma\ .$$
Then from (\ref{crules}),
\begin{eqnarray}
\eta(R_{1,j,\sigma}(\vec v)) &=& \left| \frac{v_1+v_j}{2} + \frac{|v_1-v_j|}{2}\sigma\right|^4\nonumber\\
&=&\frac{1}{8}\left| |v_1|^2+|v_j|^2 + |v_1-v_j|(v_1+v_j)\cdot \sigma\right|^2\nonumber\\
&=& \frac{1}{8}\left((|v_1|^2+|v_j|^2)^2 + 2(|v_1|^2+|v_j|^2) |v_1-v_j|(v_1+v_j)\cdot \sigma +  |v_1-v_j|^2((v_1+v_j)\cdot \sigma)^2\right)\ .\nonumber\\
\end{eqnarray}

Integrating over $S^2$ using Lemma \ref{sigmaints} yields
\begin{eqnarray}
\int_{S^2}\eta(R_{1,j,\sigma}(\vec v))b\left(\sigma\cdot \frac{v_i-v_j}{|v_i-v_j|}\right){\rm d}\sigma
&=& \frac{1}{8}(|v_1|^2+|v_j|^2)^2\nonumber\\
&+&  \frac{1}{4}(|v_1|^2+|v_j|^2)B_1(|v_1|^2-|v_j|^2) \nonumber\\
&+&    |v_1-v_j|^2|v_1+v_j|^2\frac{1-B_2}{16} + ((v_1-v_j)\cdot(v_1+v_j))^2\frac{3B_2-1}{16}\ .\nonumber\\
\end{eqnarray}
The right hand side simplifies to
\begin{eqnarray}
\frac{1}{8}(|v_1|^4+|v_j|^4 + 2|v_1|^2|v_j|^2) &+& \frac{B_1}{4}(|v_1|^4 - |v_j|^4)\nonumber\\
&+& (|v_1|^4+|v_j|^4 + 2|v_1|^2|v_j|^2 - 4(v_1\cdot v_j)^2) 
\frac{1-B_2}{16} \nonumber\\
&+&   (|v_1|^4+|v_j|^4 - 2|v_1|^2|v_j|^2)\frac{3B_2-1}{16}\ .\nonumber\\
\end{eqnarray}
It is now a simple matter to carry out the sum on $j\ge 2$. Using the identities
$$\sum_{j=2}^N|v_j|^4 = \phi(\vec v) - |v_1|^4\qquad{\rm and}\qquad  \sum_{j=2}^N|v_j|^2 = \phi(\vec v) - |v_1|^2\ ,$$
and the symmetrizing, one finds
\begin{equation}\label{tr2a}
Q\phi =  \phi -\frac{1-B_2}{N} \left[\frac{N+1}{N-1}\phi + \frac{1}{(N-1)}\psi -  \frac{2}{(N-1)}\ \right]\ .
\end{equation}
where
\begin{equation}\label{psidef}
\psi = \sum_{i \not=j} (v_i \cdot v_j)^2 \ .
\end{equation}

For $N\ge 4$, the two functions $\phi$ and $\psi$ are linearly independent, although for $N=3$ they are not. In fact for $N=3$, one has the identity
\begin{equation}
\psi = 2 \phi -\frac{1}{2} \ .
\end{equation}
Evidently, for $N\ge 4$, $\phi - \int_{X_N}\phi{\rm d}\sigma_N$ is not an eigenfunction of $Q$, so that $E_{0,2}$ is not an eigenspace of $Q$. 
We are required to compute  $Q\psi$.

We again take advantage of the permutation symmetry, and note that 
$${\cal S}((v_1\cdot v_2)^2) = \frac{2}{N(N-1)}\psi
 \qquad{\rm and}\qquad Q\psi = \frac{N(N-1)}{2}  {\cal S}(Q(v_1\cdot v_2)^2)\ .$$
We carry out the calculation in the same way that we calculated $Q\phi$, and find
\begin{equation}\label{tr1a}
Q\psi =  \psi -  \frac{1-B_2}{N} \left[     3\psi + \frac{(N-3)}{(N-1)}\phi -\frac{1}{N}\right]
\end{equation}

We see that the subspace spanned by $\phi - \int_{X_N}\phi{\rm d}\sigma_N$ 
and $\psi  - \int_{X_N}\phi{\rm d}\sigma_N$ is invariant under $Q$. 
Using (\ref{tr1a}) and (\ref{tr2a}) we easily find that the two eigenvalues of $N(I-Q)$ on the two dimensional space $V_{2,0}$ are
the eigenvalues of
$$\frac{1-B_2}{N-1}  \left[
\begin{array}{ccc}
N+1   &    1\\ 
N-3  & 3N-3\\
\end{array} \right]\ ,$$
which are
$$\frac{1-B_2}{N-1}  \left[ (2N-1) \pm \sqrt{N^2-3N +1}\right]\ .$$
The minus sign clearly gives the lesser of these, and gives the gap for $N(I-Q)$ on  $V_{2,0}$.
From here, one easily deduces (\ref{otwoq1}).

A further, much simpler calculation shows that the three functions
\begin{equation}
\psi_{1,1}^\alpha = \sum_{k=1}^N |v_k|^2 v_k^\alpha
\end{equation}
where $\alpha$ indexes the components, are also eigenfunctions of $Q$,
more precisely
\begin{equation}
Q\psi_{1,1}^\alpha = \left(1 - \frac{1-B_2}{N-1}\right)\psi_{1,1}^\alpha \ .
\end{equation}
Thus the unique eigenvalue of $A$ on $V_{1,1}$ is
$$\lambda_{1,1}^Q = 1 - \frac{1-B_2}{N-1}\ .$$

For $E_{0,2}$, a simple computation shows that the functions
\begin{equation}
\psi_{0,2}^{\alpha,\beta} = \sum_k v_k^\alpha v_k^\beta
\end{equation}
where $\alpha \not= \beta$ are indices for the components, are also  eigenfunctions for $Q$, in fact
\begin{equation}
Q\psi_{0,2}^{\alpha, \beta} = \left(1-\frac{3(1-B_2)}{2(N-1)}\right) \psi_{0,2}^{\alpha, \beta} \ .
\end{equation}  
Thus, $V_{0,2} = E_{0,2}$, and the unique eigenvalue of $Q$ on this subspace is
$$\lambda_{0,2}^Q =  1-\frac{1}{N-1}\ .$$

Finally, we consider the spectrum on $Q$ on the eigenspaces of $P$ corresponding to $n+\ell =1$. In this case, as noted above, the eigenfunctions are antisymmetric, so that if we are only concerned with the spectrum of $Q$ on the subspace of symmetric functions (which is all that is of significance for
Kac's application to the Boltzmann equation), we can ignore these eigenspaces. However,
 if we define $\eta_{0,1}(\vec v) = v_1 - v_2$ and 
$\eta_{1,0}(\vec v) = |v_1|^2 - |v_2|^2$,  we find, as above,  that
\begin{equation}\label{nlonegaps}
Q\eta_{1,0} =  \left(1-(1-B_1)\frac{1}{N-1}\right)\eta_{1,0} \qquad{\rm and}\qquad 
Q\eta_{0,1} =  \left(1-(1-B_1)\frac{1}{N-1}\right)\eta_{0,1}\ .
\end{equation}

\qed

\medskip

Now that we have all of our eigenvalues we need to order them.  By a simple comparison, we determine that  for all $N$,
the largest eigenvalue of $Q$ on our
three invariant subspaces with $n+\ell = 2$  is $\lambda_{1,1}^Q$. This is true for all choices of $b$, since the only dependence on $b$ in these eigenvalues is a common factor of $(1-B_2)$.

It is worth noting, however, that
for large $N$,
$$\lambda_{1,1}^Q = 1 - (1-B_2)\frac{1}{N} + {\cal O}\left(\frac{1}{N^2}\right) \qquad{\rm and}\qquad
\lambda_{2,0}^Q = 1 - (1-B_2)\frac{1}{N} + {\cal O}\left(\frac{1}{N^2}\right)\ ,$$
so that these eigenvalues merge as $N$ tends to infinity. Still, for all finite $N$, 
$$\lambda_{1,1}^Q = 1 - (1-B_2)\frac{1}{N-1}$$ 
is strictly larger.

Next, the invariant subspaces of $Q$ with $n+\ell =1$ are also eigenspaces of $Q$ with the eigenvalue 
$$\lambda_{0,1}^Q = \lambda_{1,0}^Q = 1- (1-B_1)\frac{1}{N-1}\ .$$

\medskip

In summary, the largest eigenvalue of $Q$ on the invariant subspaces $V_{n,\ell}$ in $L^2(X_N,\d \sigma_N)$ with $n+\ell =1,2$ and $N\ge 3$
is either
$$1 - (1-B_2)\frac{1}{N-1}\qquad{\rm or}\qquad  1- (1-B_1)\frac{1}{N-1}\ ,$$
depending on which of these is larger. 
In particular,
\begin{equation}\label{upperbnd}
\Delta_3 \le  \min\{(1-B_2),(1-B_1)\}\frac{3}{2}\ 
\end{equation}

With the above arguments we have all the ingredients needed to  prove Theorem \ref{exact}:

\medskip

\noindent{\bf Proof of Theorem \ref{exact}:}   
   First, we wish to apply Lemma \ref{mo}   to estimate $\Delta_3$ in terms of $\Delta_2$.
   In (\ref{bcn35}), we have set $\mu_3^\star = 11/20$, and with this choice of the threshold, we have seen that there is just
   one eigenvalue of $P$ between $\mu_3^\star $ and $1$, namely the gap eigenvalue $\mu_3 = \mu_{1,1}(3) =  \mu_{0,1}(3) = 
   \mu_{1,0}(3)$.
   Thus, from   Lemma \ref{mo} and the eigenvalue computations in Lemma \ref{lemQ}, either the gap eigenvalue of $Q$ for $N=3$ is
   \begin{equation}\label{bcn70}
   \max\left\{\  1 - (1-B_2)\frac{1}{N-1}\ ,\  1- (1-B_1)\frac{1}{N-1}\ \right\}\ ,
   \end{equation}
   or else
   \begin{equation}\label{bcn71}
   \Delta_3\ge \frac{3}{2} \left( 1- \frac{11}{20}\right)\Delta_2 = \frac{27}{40}\Delta_2\ .
   \end{equation}
   If (\ref{bcn70}) does give the gap eigenvalue of $Q$ for $N=3$, then 
   \begin{equation}\label{bcn72}
   \Delta_3 = \min\{\ (1-B_1)\ ,\ (1-B_2)\ \}\frac{3}{2}\ .
   \end{equation}

   Since according to Lemma \ref{mo}, at least one of (\ref{bcn71}) and  (\ref{bcn72}) is true, the condition
   \begin{equation}\label{d2cond4}
\frac{27}{40}\Delta_2 \ge  \frac{3}{2} \min\{\ (1-B_1)\ ,\ (1-B_2)\ \}\ ,
\end{equation}
and (\ref{upperbnd}) ensure that   (\ref{bcn72}) is true, and thus gives us the gap eigenvalue for $N =3$. 
Note that the condition (\ref{d2cond4}) is equivalent to the condition  (\ref{d2cond}) in Theorem \ref{exact}.

   Now we proceed by induction. For any $n\ge 4$, assume that   
     \begin{equation}\label{bcn73}
   \Delta_{N-1} = \min\{\ (1-B_1)\ ,\ (1-B_2)\ \}\frac{N-1}{N-2}\ .
   \end{equation}
   
   In (\ref{bcn34}) we have set
   $$\mu_N^\star = \frac{1}{N-1}$$
   for all $N\ge 4$, and we have seen that the only eigenvalues $\mu$ of $P$  with $1 > \mu \ge  \mu_N^\star$
   are the gap eigenvalue   $\mu_N = \mu_{1,1}(N) =  \mu_{0,1}(N) = 
   \mu_{1,0}(N)$, and for $N>4$,   $\mu_{2,0}(N)$. 
   Thus, by Lemma \ref{mo} and the eigenvalue computations in Lemma \ref{lemQ}, either the gap eigenvalue of $Q$ for $N$ is
   \begin{equation}\label{bcn75}
   \max\left\{\  1 - (1-B_2)\frac{1}{N-1}\ ,\  1- (1-B_1)\frac{1}{N-1}\ \right\}\ ,
   \end{equation}
   or else
   \begin{equation}\label{bcn76}
   \Delta_N >  \frac{N}{N-1} \left( 1- \frac{1}{N-1}\right)\Delta_{N-1}\ .
   \end{equation}
   There is strict inequality in (\ref{bcn76}) since all remaining eigenvalues of $P$ not taken into account in (\ref{bcn75}) are strictly less than 
   $\mu_N^\star$. 
   By the inductive hypothesis (\ref{bcn73}),  (\ref{bcn73}) yields
   $$\Delta_N >   \min\{\ (1-B_1)\ ,\ (1-B_2)\ \}\frac{N}{N-1}\ .$$
   This is impossible,  as the trial functions leading to (\ref{bcn75}) yield the upper bound
   \begin{equation}\label{bcn77}
   \Delta_N \le   \min\{\ (1-B_1)\ ,\ (1-B_2)\ \}\frac{N}{N-1}\ .
   \end{equation}
   Thus equality holds in (\ref{bcn77}), which completes the proof of the inductive step.
    Because of the strict inequality in (\ref{bcn76}), the only eigenfunctions with the gap eigenvalue are found in the invariant subspaces
    considered here; i.e., in the $V_{n,\ell}$ with $0 < n+\ell \le 2$. By the results of Lemma \ref{lemQ}, this yields the statement in Theorem \ref{exact} concerning the  gap eigenfunctions of $Q$. \qed
    
   \medskip

\medskip

\noindent{\bf Proof of Theorem \ref{exact2}:}   We proceed as in the previous proof, except that for low values of $N$, we shall use a different choice for the threhold $\mu_N^\star$, namely
\begin{equation}\label{fin2}
\mu_N^\star = \mu_{2,2}(N)\ .
\end{equation}
We know from Lemma \ref{twotwo} that for all $N \le 7$, 
$$\mu_{n,\ell}(N) \le \mu_{2,2}(N)\quad{\rm for\ all}\quad n+\ell > 2\ .$$
Thus, at least for such $N$, the only eigenvalues $\mu$ of $P$ with $\mu > 
\mu_N^\star = \mu_{2,2}(N)$ are those with $n+\ell \le 2$. 
 We have already computed the gap for $Q$ on the invariant subspaces containing these
 eigenvalues, and we have found that the gap in these subspaces is
 $$\widetilde \Delta_N = \min\{\ (1-B_1)\ ,\ (1-B_2)\ \}\frac{N}{N-1}\ .$$
 
 If for any $N_0\ge 3$, it turns out $\widetilde \Delta_{N_0} = \Delta_{N_0}$, the gap on the whole space, then
 we can switch from that point onwards to the use of  $\mu_N^\star = \mu_{0,2}(N)$ as in the proof of Theorem \ref{exact} to show that  $\widetilde \Delta_{N} = \Delta_{N}$ for all $N\ge N_0$, and that the eigenfunctions are exactly as claimed for all $N > N_0$.
 
We now show that it is always the case that $\widetilde \Delta_{N_0} = \Delta_{N_0}$
 for some $N_0 \le 7$.  To do this, pick any value $N_1 \ge 4$, and suppose that for $3 \le j \le N_1$, we have
 \begin{equation}\label{fin3}
 \Delta_j < \min\{\ (1-B_1)\ ,\ (1-B_2)\ \}\frac{j}{j-1}\ .
 \end{equation}

Then by Lemma \ref{mo} and  Lemma \ref{twotwo}, using the value 
$\mu_j^\star = \mu_{2,2}(j)$, for $3\le j \le N_1$, we have
$$\Delta_{N_1} \ge \frac{N_1}{2}\prod_{j=3}^{N_1}(1- \mu_{2,2}(j))\Delta_2\ .$$
Using the hypothesis $\Delta_2 = 2(1- B_1)$, we have
$$\Delta_{N_1} \ge \frac{N_1}{2}\prod_{j=3}^{N_1}(1- \mu_{2,2}(j))2(1-B_1)\ .$$

Of course we can rewrite this as
\begin{eqnarray}\Delta_{N_1} &\ge& \frac{N_1}{2}\left(\prod_{j=4}^{N_1}(1- \mu_{0,2}(j))\right)
(1- \mu_{2,2}(3))\left(\prod_{j=4}^{N_1}
\frac{(1- \mu_{2,2}(j)}{(1- \mu_{0,2}(j)}\right)
2(1-B_1)\nonumber\\
&=& \frac{N_1}{N_1-1}\left(\prod_{j=4}^{N_1}
\frac{(1- \mu_{2,2}(j)}{(1- \mu_{0,2}(j)}\right)
\frac{9}{10}(1-B_1)\ ,\nonumber\\
\end{eqnarray}
since, as in the last proof,  $(1- \mu_{2,2}(3)) = 9/20$, and 
$$ \frac{N_1}{2}\prod_{j=4}^{N_1}(1- \mu_{0,2}(j)) = \frac{N_1}{N_1-1}\ .$$
Now, by direct computation, we find
$$\prod_{j=4}^{7}
\frac{(1- \mu_{2,2}(j)}{(1- \mu_{0,2}(j)} = \frac{558018643}{495720000}
> \frac{10}{9}\ .$$
For $N_1\ge 7$, this would lead to 
$$\Delta_{N_1} > \frac{N_1}{N_1-1}(1-B_1)\ ,$$
and this is impossible, since we have a trial function showing that the gap cannot be so large. Hence it must be that
(\ref{fin3}) is false for some $j\le 7$. By what we have said above, from this point onward, we can proceed as in the proof of Theorem \ref{exact}, and we obtain Theorem \ref{exact2}
\qed

\medskip

While the results presented here cover a very wide range of models, it is possible to come up with choices of $b$ for which $\Delta_2 \ne 2(1-B_1)$. If one found a need  to deal with such an example, one might have to go deeper into the spectrum of $P$.
It is very likely that Lemma \ref{twotwo} holds for all $N\ge 3$, based on extensive computation. These computations also show that
as $N$ increases, $\mu_{2,1}(N)$ comes very close to  $\mu_{2,2}(N)$, so that to get much more leverage,
 one would need to  compute all of the eigenvalues of $Q$ on the smallest invariant subspaces of $Q$ that contains both of these eigenspaces of $P$.  This could be done using the methods presented here, but the computations would be considerably more involved than the ones we have presented in this section. 
Thus, having treated a wide range of models, we shall
conclude our discussion of $Q$ here. In the brief final section, we discuss a point we raised earlier concerning bounds on Jacobi polynomials.

\medskip
\section{Bounds on Jacobi polynomials}
\medskip

As alluded to in Section 8 the identity (\ref{bcnra}), together with the trivial bound on the $|\kappa_{n,\ell}|\le 1$, which comes from the
fact that $K$ is a Markov operator, 
will for certain regions provide a stronger bound than  (\ref{nembnd}), the bound of Nevai, Erdelyi and Magnus.  We close this section by showing how   (\ref{bcnra}) can be used to obtain better bounds.

To begin, write
\begin{equation}
b^{2\beta-1}\frac{P_n^{\alpha,\beta}(-1+2b^2)}{P_n^{\alpha,\beta}(1)}\le\frac{2e}{\pi}\frac{\Gamma(n+1)}{b(1-b^2)^{\alpha+1/2}}\frac{2+\sqrt{\alpha^2+\beta^2}}{2n+\alpha+\beta+1}\frac{\Gamma(n+\beta+1)}{\Gamma(n+\alpha+\beta+1)}\frac{\Gamma(\alpha+1)^2}{\Gamma(n+\alpha+1)},
\end{equation}
where $\beta=l+1/2$ with $l$ an integer. In regions where the right hand side of the above equation becomes larger than one the simple bound 
$$b^{2\beta-1}\frac{P_n^{\alpha,\beta}(-1+2b^2)}{P_n^{\alpha,\beta}(1)}\le1$$ becomes stronger. In the region $2n+1<\alpha<\beta$, we find $\frac{2+\sqrt{\alpha^2+\beta^2}}{2n+\alpha+\beta+1}>\frac{1}{4}$. This plus Stirling's formula with the remainder yields,
\begin{align*}
&\frac{2e}{\pi}\frac{\Gamma(n+1)}{b(1-b^2)^{\alpha+1/2}}\frac{2+\sqrt{\alpha^2+\beta^2}}{2n+\alpha+\beta+1}\frac{\Gamma(n+\beta+1)}{\Gamma(n+\alpha+\beta+1)}\frac{\Gamma(\alpha+1)^2}{\Gamma(n+\alpha+1)}\\& >\frac{e^n}{\sqrt{2\pi}}\frac{\Gamma(n+1)}{b(1-b^2)^{\alpha+1/2}}\frac{\alpha^{\alpha+1/2-n}\beta^{-\alpha}}{(1+\frac{\alpha}{\beta})^{n+\alpha+\beta+1}}\frac{(1+\frac{1}{\alpha})^{2\alpha+1}}{(1+\frac{n+1}{\alpha})^{n+\alpha+1/2}(1+\frac{n+1}{\alpha+\beta})^{\beta}}*r\\&>\frac{e^n}{\sqrt{2\pi}}\frac{\Gamma(n+1)}{b(1-b^2)^{\alpha+1/2}}\frac{\alpha^{\alpha+1/2-n}\beta^{-\alpha}}{2^{2n+2\alpha+2\beta+3/2}}r,
\end{align*}
where ${\displaystyle r=\left(1-\frac{1}{12(n+\alpha+\beta+1)}\right)\left(1-\frac{1}{12(n+\alpha+1)}\right)}$ and $n$ is assumed to be fixed. Choosing $b(1-b^2)^{\alpha+1/2}$ so that the last inequality is greater than one provides a region where (\ref{bcnra}) and  $|\kappa_{n,\ell}|\le 1$ does better than
 (\ref{bcnra}). It would be interesting to obtain better bounds on   $|\kappa_{n,\ell}|$ by direct analysis of $K$, and to use these to sharpen
 the argument just made.
\medskip

\noindent{\bf Acknowledgement}  We would like to thank Doron Lubinsky for valuable discussions concerning
 (\ref{nembnd}), the bound of Nevai, Erdelyi and Magnus, and related results.

\end{document}